\documentclass[
  aps,
  prb,
  reprint,
  superscriptaddress,
  longbibliography,
  showkeys,
  citeautoscript,
]{revtex4-1}

\usepackage[intlimits]{amsmath}
\usepackage{amssymb}
\usepackage{amsthm}
\usepackage{booktabs}
\usepackage{braket}
\usepackage{color}
\usepackage{float}
\usepackage{graphicx}
\usepackage{glossaries}
\usepackage[colorlinks=true,
            linkcolor=NavyBlue,
            citecolor=NavyBlue,
            filecolor=NavyBlue,
            urlcolor=NavyBlue
            ]{hyperref}
\usepackage[utf8]{inputenc}
\usepackage[version=4]{mhchem}
\usepackage{multirow}
\usepackage{soul} 
\usepackage{tabularx}
\usepackage{units}
\usepackage[dvipsnames]{xcolor}

\setacronymstyle{long-short}
\newacronym{tmd}{TMD}{transition metal dichalcogenide}
\newacronym{pdms}{PDMS}{polydimethylsiloxane}
\newacronym{vdw}{vdW}{van der Waals}
\newacronym{dft}{DFT}{density functional theory}
\newacronym{vis}{VIS}{visible}
\newacronym{nir}{NIR}{near-infrared}
\newacronym{uv}{UV}{ultraviolet}
\newacronym{mse}{MSE}{mean squared error}
\newacronym{si}{SI}{Supporting Information}

\begin{document}

\title{Optical constants of several multilayer transition metal dichalcogenides measured by spectroscopic ellipsometry in the 300-1700 nm range: high-index, anisotropy, and hyperbolicity}

\newcommand{\phys}{
    Department of Physics,
    Chalmers University of Technology,
    SE-412~96 Gothenburg, Sweden
}
\newcommand{\dtu}{
    Department of Photonics Engineering, 
    Technical University of Denmark, 
    2800 Kongens Lyngby, Denmark
}
\newcommand{\warsaw}{
    Faculty of Physics,
    University of Warsaw,
    Pasteura 5,
    PL-02-093 Warsaw, Poland
}

\author{Battulga Munkhbat}
\altaffiliation{Contributed equally to this work}
\affiliation{\phys}
\affiliation{\dtu}

\author{Piotr Wr\'obel}
\altaffiliation{Contributed equally to this work}
\affiliation{\warsaw}

\author{Tomasz J. Antosiewicz}
\email[]{tomasz.antosiewicz@fuw.edu.pl}
\affiliation{\warsaw}
\affiliation{\phys}

\author{Timur O. Shegai}
\email[]{timurs@chalmers.se}
\affiliation{\phys}

\begin{abstract}
Transition metal dichalcogenides (TMDs) attract significant attention due to their exceptional optical and excitonic properties. It was understood already in the 1960s, and recently rediscovered, that many TMDs possess high refractive index and optical anisotropy, which make them attractive for nanophotonic applications. However, accurate analysis and predictions of nanooptical phenomena require knowledge of dielectric constants along both in- and out-of-plane directions and over a broad spectral range -- information, which is often inaccessible or incomplete. Here, we present an experimental study of optical constants from several exfoliated TMD multilayers obtained using spectroscopic ellipsometry in the broad range of 300--1700 nm. The specific materials studied include semiconducting \ce{WS2}, \ce{WSe2}, \ce{MoS2}, \ce{MoSe2}, \ce{MoTe2}, as well as, in-plane anisotropic \ce{ReS2}, \ce{WTe2}, and metallic \ce{TaS2}, \ce{TaSe2}, and \ce{NbSe2}. The extracted parameters demonstrate high-index ($n$ up till $\approx 4.84$ for \ce{MoTe2}), significant anisotropy ($n_{\parallel}-n_{\perp} \approx 1.54$ for \ce{MoTe2}), and low absorption in the near infrared region. Moreover, metallic TMDs show potential for combined plasmonic-dielectric behavior and hyperbolicity, as their plasma frequency occurs at around $\sim$1000--1300 nm depending on the material. The knowledge of optical constants of these materials opens new experimental and computational possibilities for further development of all-TMD nanophotonics. 
\end{abstract}

\maketitle

\section{Introduction}

The recent interest in 2D semiconductors stems from the direct band gap of monolayer \ce{MoS2} \cite{splendiani2010emerging, mak2010atomically}. In addition to exciting monolayer physics, multilayer \glspl{tmd} possess a number of attractive optical properties. The prospects of multilayer \glspl{tmd} for optics have been discussed already in the 1960s \cite{wilson1969transition}. For example, due to their \gls{vdw} nature, \glspl{tmd} are naturally anisotropic, which is reflected in their physical and optical properties \cite{fei2016nano, hu2017imaging, hu2017probing, babicheva2018near, ermolaev2020giant}. Due to large oscillator strengths of electronic excitations around the A-, B- and C-exciton bands, which are stable in both mono- and multilayer forms and even at room temperature, \gls{tmd} materials possess high refractive indexes in the visible and near infrared range \cite{li2014measurement, ermolaev2020broadband, li2014broadband, yim2014investigation}. Moreover, below the A-exciton absorption band, there is a relatively broad region of low loss \cite{hu2017probing, ermolaev2020broadband}. 

These observations have led to a recently renewed interest in \gls{tmd} optics and nanophotonics. This interest has grown even more after a publication of several nanopatterning methods of \glspl{tmd} and studies of optical phenomena in resulting \gls{tmd} nanostructures \cite{verre2019transition, zhang2019guiding}. 
This includes recent observations of high-index Mie resonances and anapole states in \ce{WS2} nanodisks \cite{verre2019transition}, optical anisotropy in \gls{tmd} slabs \cite{hu2017probing, ermolaev2020giant} and nanocones \cite{green2020optical}, self-hybridization in \gls{tmd} slabs \cite{munkhbat2018self} and nanotubes \cite{yadgarov2018strong}, optical modes in lattices of \gls{tmd} nanostructures \cite{babicheva2019lattice}, improved second-harmonic generation in \ce{WS2} and \ce{MoS2} disks \cite{busschaert2020tmdc,nauman2021tunable}, high-index metamaterials \cite{krishnamoorthy2020infrared}, nanoholes down to $\sim$20 nm \cite{danielsen2021super}, and \gls{tmd} metamaterials with atomically sharp edges \cite{munkhbat2020transition}.

Theoretical predictions of large values of dielectric functions of \glspl{tmd} and related \gls{vdw} materials include \gls{dft} studies, which confirm exceptionally high and anisotropic permittivity tensors in these materials \cite{laturia2018dielectric, shubnic2020high}. Theoretically, \ce{MoTe2}, \ce{ReS2}, and ReSe$_2$ compounds were predicted to have the highest values of the in-plane dielectric function. Together, these experimental observations and theoretical predictions strongly motivate further development of \gls{tmd} nanophotonics. The knowledge of optical constants of these materials is essential in this regard. However, these parameters are often not precisely known, the out-of-plane values are often assumed, the response is measured within too narrow a spectral range. In addition, the quality of the extracted optical constants for a \gls{tmd} material highly depends on the sample preparation \cite{cai2018chemical, ermolaev2020broadband}, number of layers \cite{diware2017characterization,kylanpaa2015binding}, and their lateral dimensions. The emerging field of \gls{tmd} nanophotonics needs a reliable and accurate library of both in-plane and out-of-plane optical constants of various \glspl{tmd} over a broad spectral range to explore their potential not only in the visible but also in the near-infrared optoelectronic applications. Here, we present such a library of \glspl{tmd} which have been characterized using the same very sensitive and accurate ellipsometric approach which accounts for sample and  measurement nonidealities such as surface roughness, thickness non-uniformity, as well as spectrometer bandwidth and beam angular spread, ensuring a good level of consistency between all our results.

Specifically, we have measured both in-plane and out-of-plane optical constants of high-quality mechanically-exfoliated \glspl{tmd} over a broad spectral range of 300--1700 nm using spectroscopic ellipsometry. Specifically, we investigated several important multilayer \glspl{tmd} with large lateral dimensions, including semiconducting \ce{WS2}, \ce{WSe2}, \ce{MoS2}, \ce{MoSe2}, \ce{MoTe2}, as well as, in-plane anisotropic \ce{ReS2}, \ce{WTe2}, and metallic \ce{TaS2}, \ce{TaSe2}, and \ce{NbSe2}. Our experimentally obtained data from such high-quality \glspl{tmd} reveal several interesting optical properties that are promising for future nanophotonic and optoelectronic applications.

\begin{figure}
\includegraphics[width=8.5 cm]{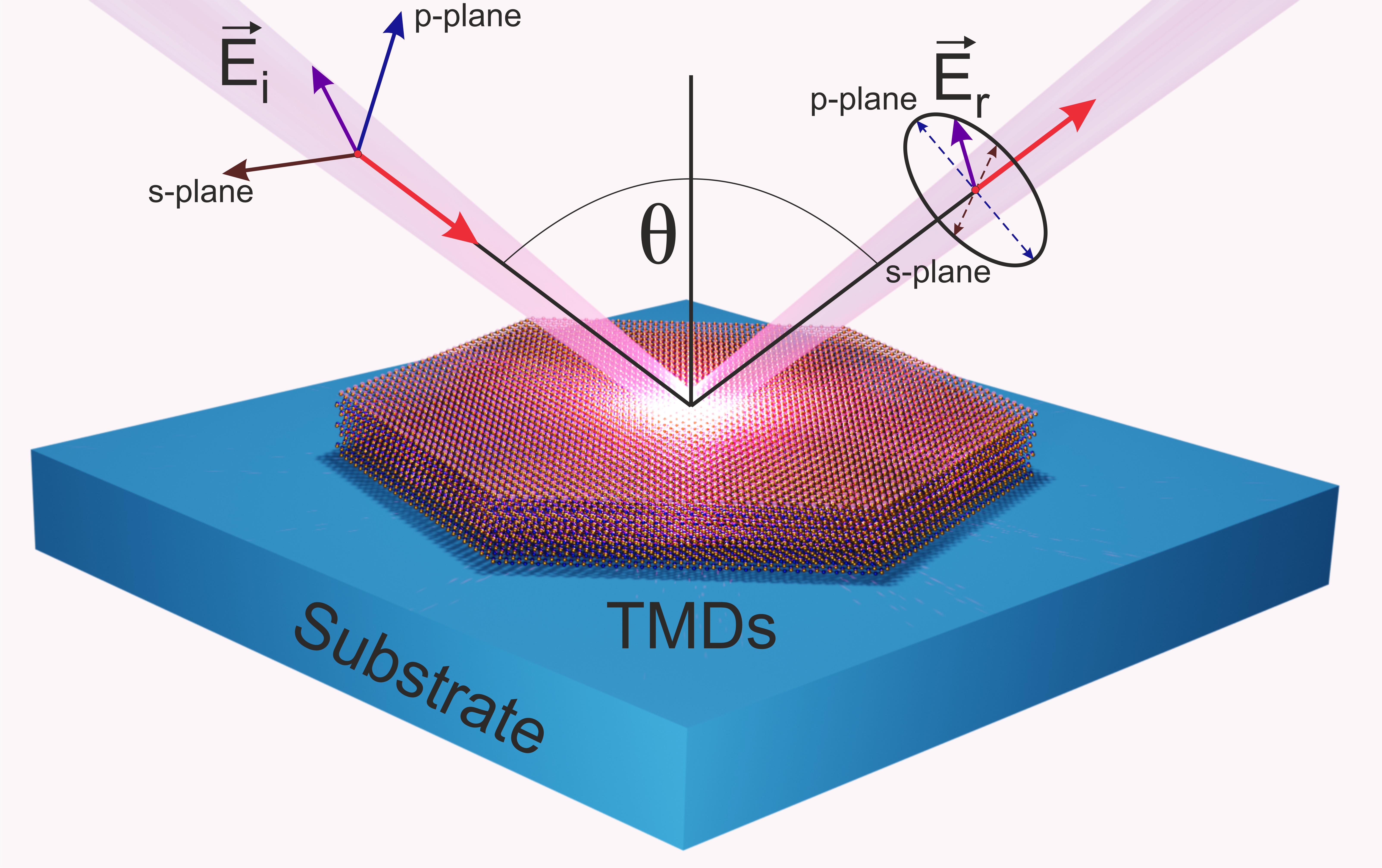}
\caption{Schematic of the spectroscopic ellipsometry measurements on mechanically exfoliated multilayer \glspl{tmd} on a substrate.
}
\label{fig:sketch}
\end{figure}  

\section{Results and discussion}
To extract the optical constants, we prepared all multilayer \gls{tmd} flakes by mechanical exfoliation directly from high quality bulk crystals (HQ-graphene) onto substrates. However, in order to perform spectroscopic ellipsometry measurements with low uncertainty, the lateral dimensions of the \gls{tmd} flakes are critical, and must be larger than the beam spot during the measurements.

\begin{figure}[ht]
\includegraphics[width=6.7cm]{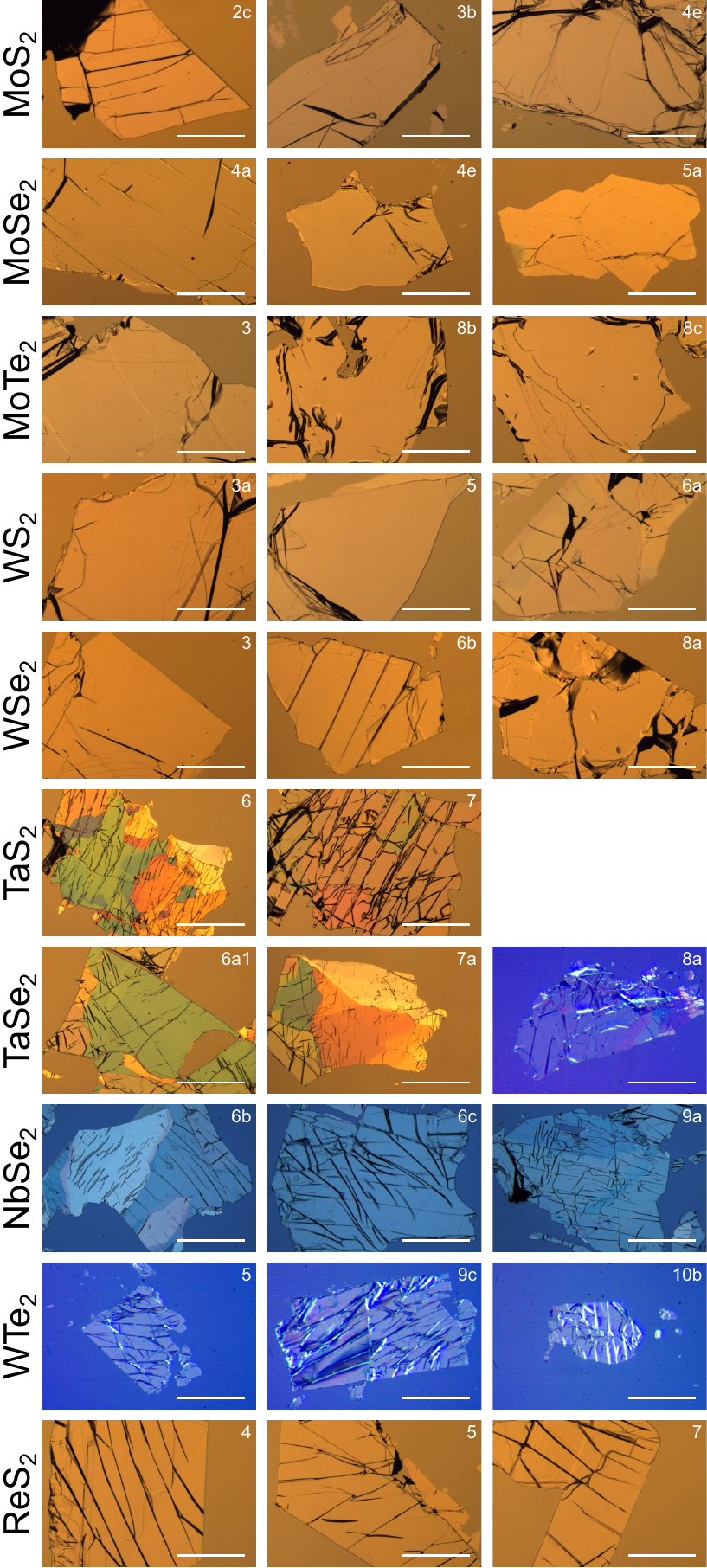}
\caption{Exemplary images of exfoliated \gls{tmd} flakes used ellipsometric measurements. First five rows show uniaxial semiconducting \gls{tmd} flakes: \ce{MoS2}, \ce{MoSe2}, (2H)\ce{MoTe2}, \ce{WS2}, and \ce{WSe2},  respectively. Next three rows contain uniaxial metallic \gls{tmd} flakes: \ce{TaS2}, \ce{TaSe2}, and \ce{NbSe2}. The last two rows contain the two characterised biaxial \gls{tmd} flakes: metallic (though in the optical range the permittivity is positive) \ce{WTe2} and semiconducting \ce{ReS2}, respectively. All semiconducting \gls{tmd} flakes were placed directly on a Si substrate with a rough (diffusive) backside. In the case of metallic \glspl{tmd} a thermally oxidized Si substrate was used (with \ce{SiO2} thickness of \unit[3]{$\mu$m} or \unit[8.8]{$\mu$m}). The scale bar is \unit[300]{$\mu$m} long and the same in every panel.
}
\label{fig:flakes}
\end{figure}

In this work, the spectroscopic ellipsometry measurements, see \autoref{fig:sketch} for scheme, were performed using a variable angle spectroscopic ellipsometer with a Dual Rotating Compensator design (VASE Woollam RC2) equipped with focusing probes to reduce the beam diameter to $\sim$300~$\mu$m. To provide multilayer \glspl{tmd} of sufficient lateral dimensions, all multilayer \gls{tmd} flakes were mechanically exfoliated from bulk crystals, first, onto \gls{pdms} stamps using the scotch-tape method. Subsequently, the partially transparent semiconducting flakes were transferred onto one-side polished silicon substrates with a self-limiting natural oxide layer ($\sim$1-3 nm) using the all-dry-transfer method \cite{Castellanos2014transfer} with a few important concerns \cite{kinoshita2019dry}. The lossy and/or metallic \gls{tmd} flakes were transferred onto silicon substrates with thermally grown \ce{SiO2} with nominal thicknesses of \unit[3]{$\mu$m} or \unit[8.8]{$\mu$m}. First, we chose the original bulk crystals carefully, and exfoliated multilayers onto \gls{pdms} stamps only from large (at least a centimeter) homogeneous crystals using a blue scotch-tape. Second, due to the thermoplastic properties of the \gls{pdms} film, the adhesion between the \gls{tmd} flakes and \gls{pdms} slightly decreases at elevated temperature (here 60$^{\circ}$C). By exploiting this property, large multilayer \glspl{tmd} with relatively homogeneous thickness can be readily transferred onto a substrate for ellipsometric measurements. Thicknesses of the transferred \gls{tmd} flakes were measured using a VEECO profilometer. For our study, we chose multilayer \glspl{tmd} with thicknesses ranging from a few tens of nanometers to several microns. Exemplary \gls{tmd} flakes are shown in \autoref{fig:flakes}. These include semiconducting \ce{WS2}, \ce{WSe2}, \ce{MoS2}, \ce{MoSe2}, \ce{MoTe2}, as well as, bianisotropic \ce{ReS2}, \ce{WTe2}, and metallic \ce{TaS2}, \ce{TaSe2}, and \ce{NbSe2}, which are among the most promising \glspl{tmd} for future nanophotonic and optoelectronic applications. After sample preparation, we performed spectroscopic ellipsometry measurements and analysis for all freshly prepared multilayers in a broad spectral range of 300--1700 nm in steps of 1 nm. The measurements were performed at multiple angles of incidence ranging from 20$^{\circ}$ to 75$^{\circ}$ in steps of 5$^{\circ}$, although for some flakes the maximum angle was reduced to ensure that the illumination spot was smaller than a \gls{tmd} flake. The obtained dielectric tensor components data as a function of wavelength for all studied materials are provided in the \gls{si}.

\subsection{Uniaxial semiconductors}
We begin our study by investigating multilayers of uniaxial \gls{tmd} semiconductors, such as \ce{MoS2}, \ce{MoSe2}, \ce{MoTe2}, \ce{WS2}, and \ce{WSe2}. These are one of the most well-studied \gls{tmd} materials, especially in relation to their monolayers \cite{mak2010atomically} and \gls{vdw} heterostructures \cite{rivera2018interlayer}.
Recently, they have been also suggested as promising high-index dielectric material platform for future nanophotonics \cite{li2014measurement, liu2018ultrathin, verre2019transition, shubnic2020high, ling2021all}. Here, we report both their in-plane and out-of-plane dielectric constants and compare our results to previously reported parameters.\cite{beal1976kramers, beal1979kramers, li2014measurement, ermolaev2020giant, shubnic2020high, laturia2018dielectric} Their permittivities (both real and imaginary parts) were measured utilizing the ellipsometry technique whose basic principle is depicted in \autoref{fig:sketch}. In general, ellipsometry measures changes of the polarization state of light upon reflection of an incident beam from a sample. The change is represented by two measured parameters $\Psi$ (Psi) and $\Delta$ (Delta), which correspond to the ratio of the reflection coefficients and the phase difference between the $p-$ and $s-$polarization components of the incident beam.
The described above approach assumes that no change of polarization state occurs upon reflection from the sample. In the case of anisotropic samples, cross-polarization might occur, thus more complex analysis including General Ellipsometry or Mueller Matrix ellipsometry are required to address this issue. However, in the case of uniaxial anisotropic materials proper sample alignment results in canceling all the off-diagonal components of the Mueller Matrix which simplifies the procedure to the standard ellipsometry. While this alignment is straighforward for \gls{tmd} flakes due to their \gls{vdw} nature, which dictates the alignment of the crystalline axes, the full Mueller Matrix was measured to ensure this proper alignment (\autoref{fig:mm-mos2}--\autoref{fig:mm-tase2}).

To extract physical parameters like thickness or complex refractive index of a given material, an appropriate model describing the investigated structure has to be built. Parameters of interest are extracted by a simultaneous fit of the model parameters to the Psi and Delta curves. Although the technique allows for optical characterization of a sample with thickness down to a monolayer, measurements of the anisotropic samples are challenging \cite{xu2021optical} and require thick samples to assure sufficient light interaction with in-plane and, especially, out-of-plane polarization components to sense the anisotropy. That is particularly difficult for samples with a high refractive index what reduces greatly the angle of refraction and thus the amount of the out-of-plane components of the refracted beam in the sample. This problem can be partially overcome by using an appropriate scheme of measurements and analysis that allows improving the sensitivity of the model (see Methods).

\begin{figure}
\includegraphics[width=60mm]{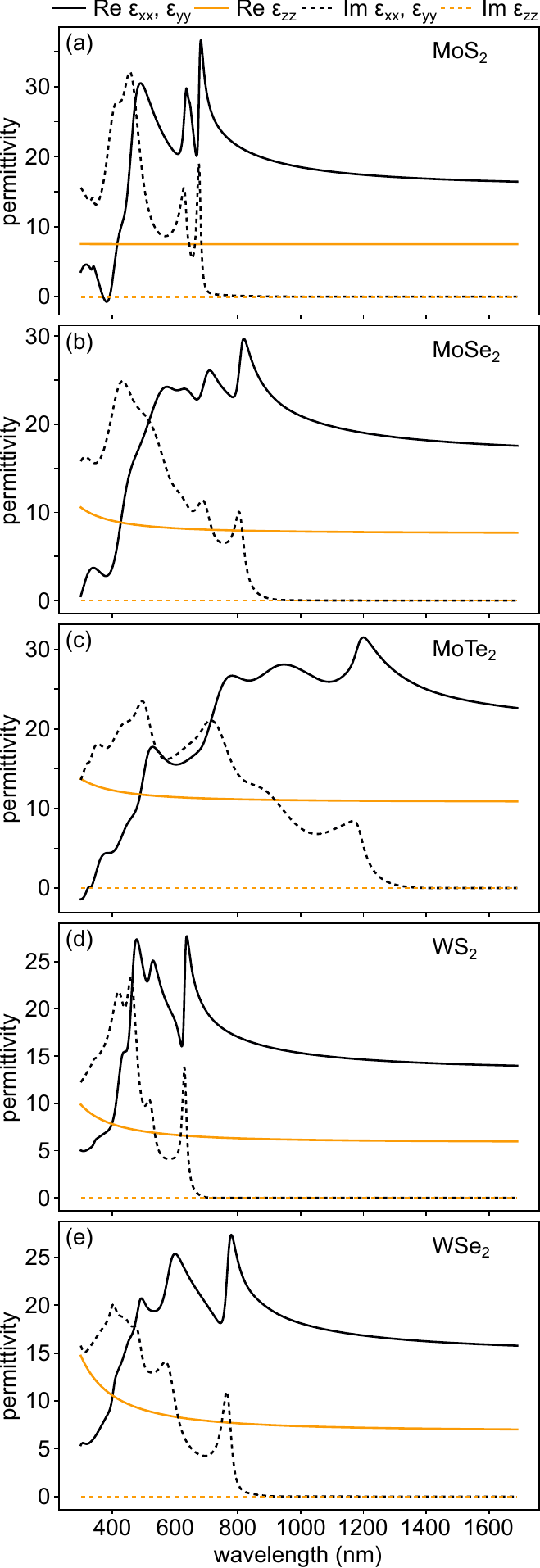}
\caption{Permittivity of uniaxial semiconducting \gls{tmd} flakes: (a) \ce{MoS2}, (b) \ce{MoSe2}, (c) \ce{MoTe2}, (d) \ce{WS2}, and (e) \ce{WSe2}. Corresponding exemplary Mueller Matrix measurements proving their uniaxial nature are plotted in \autoref{fig:mm-mos2}--\autoref{fig:mm-wse2}.
}
\label{fig:semi-uni}
\end{figure}

In our experiment, the samples exhibiting transparency in the visible-near infrared (VIS-NIR) range are in the form of thick \glspl{tmd} layers (from a few tens of nm to a few microns) exfoliated onto roughened Si substrates with a native \ce{SiO2} layer. In the extraction procedure, we use a multisample analysis approach with the model containing a semi-infinite Si substrate with a native oxide and thin layer of a \glspl{tmd}. The optical constants of Si and \ce{SiO2} are taken from the CompleteEASE database and their validity was confirmed by reference measurements of substrates next to the \gls{tmd} flakes. The in-plane component of the complex refractive index is described by multiple Tauc-Lorentz dispersion model terms, while the out-of-plane component is described by a single pole described by $\varepsilon_{\infty}$ and its \gls{uv} position and amplitude (see Methods). In the analysis, both surface roughness and layer nonuniformity are taken into account and the goodness of fit parameter, defined as the \gls{mse}, is minimized during the fitting procedure. The sensitivity of the technique to the anisotropic properties of the samples can be deduced from the asymmetry of the interference maxima in the $\Psi$ curves occurring in the transparent regions of the samples. Moreover, a significant drop of \gls{mse} when the permittivity model is changed from isotropic to anisotropic indicates that this approach allows for extracting the out-of-plane component. \autoref{fig:semi-uni} shows the complex permittivities of the analyzed uniaxial semitransparent \glspl{tmd} (see \autoref{fig:fit-mos2}--\autoref{fig:fit-wse2} for quality of the fits to $\Psi$ and $\Delta$). The surface of some of the samples is nonuniform, cf. \autoref{fig:flakes}, with wrinkles/folds and atomic steps that cannot be easily subsumed into roughness within the model. Since the surface features influence the \gls{uv} region the most, to minimize the uncertainty of the extracted dispersion curves we consider the data from \unit[300]{nm}.

\begin{figure}
\includegraphics[width=0.8\columnwidth]{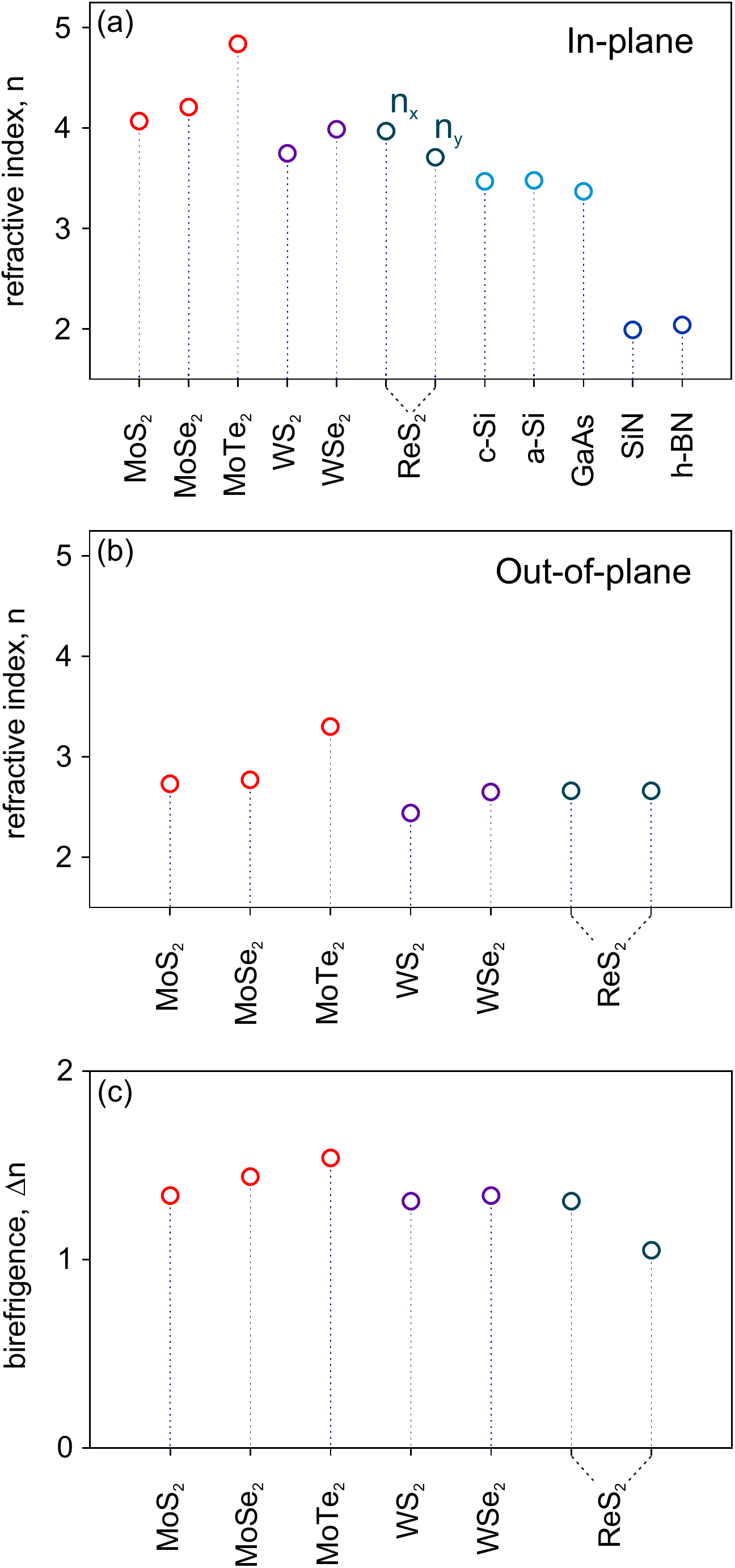}
\caption{Comparison of extracted refractive indices of common \glspl{tmd} at the Telecom C band (1550 nm) with conventional semiconductors. (a) In-plane refractive index, (b) out-of-plane refractive index, (c) birefringence.
}
\label{fig:n}
\end{figure} 

Extracted in-plane ($n_{\parallel}$, here $n_\parallel=\sqrt{\varepsilon_{xx}}=\sqrt{\varepsilon_{yy}}$) and out-of-plane ($n_{\perp}=\sqrt{\varepsilon_{zz}}$) refractive indices at \unit[1550]{nm} for 
semiconducting \glspl{tmd} and their optical anisotropy ($\Delta n$)  are displayed in \autoref{fig:n}. Overall, multilayer \glspl{tmd} exhibit higher refractive indices at 1550 nm, in contrast to conventional semiconductors, e.g., c-Si ($\sim$3.47),\cite{li1980refractive} a-Si ($\sim$3.48),\cite{pierce1972electronic} and GaAs ($\sim$3.37).\cite{papatryfonos2021refractive}
Our data in \autoref{fig:n}a reveal a few interesting trends. First, we observe higher index values among Mo-based \glspl{tmd}, $\sim$4.07 (\ce{MoS2}), $\sim$4.21 (\ce{MoSe2}), and $\sim$4.84 (\ce{MoTe2}), when compared to other W-based \ce{WS2} ($\sim$3.75), and \ce{WSe2} ($\sim$3.99) semiconducting \glspl{tmd}. 
Second, these data also reveal that the refractive index increases depending on the chalcogen atom of \glspl{tmd} in the following order: $n_{\ce{MoS2}} < n_{\ce{MoSe2}} < n_{\ce{MoTe2}}$. This observation agrees well with theoretically predicted results from an earlier \gls{dft} study.\cite{laturia2018dielectric} In parallel to high index, it is worth mentioning that their optical loss in the near-infrared range is negligibly small judging from their very small imaginary part of the permittivities, cf. \autoref{fig:semi-uni}. This property opens a possibility of using multilayer \glspl{tmd} for low-loss nanophotonics applications \cite{ling2021all, munkhbat2022nanostructured}. 

Another interesting optical feature in multilayer \glspl{tmd} is their anisotropic properties due to \gls{vdw} stacking nature, which results in large birefringence. \autoref{fig:n}c shows that \glspl{tmd} exhibit relatively large birefringence of $\Delta n\geq 1.3$ due to their lower out-of-plane refractive indices ranging from $n_{\perp}$: 2.44 (\ce{WS2}) to 3.3 (\ce{MoTe2}), in contrast to their in-plane indices (cf. \autoref{fig:n}ab). 
\ce{MoTe2} shows the largest birefringence of ${\Delta n}\sim1.54$ among uniaxial semiconducting \glspl{tmd}, whereas other materials \ce{MoS2}, \ce{MoSe2}, \ce{WS2}, and \ce{WSe2} exhibit birefringence values of ${\Delta n}\sim$1.34, $\sim$1.44, $\sim$1.31, and $\sim$1.34, respectively. 
It should be noted that the obtained birefrigence values in \glspl{tmd} are 7-8 times larger than other anisotropic materials, e.g., yttrium orthovanadate and rutile \ce{TiO2}, which exhibit typically smaller anisotropy of ${\Delta}n \sim0.2-0.3$.\cite{takayama2008dyakonov} Our data are in a good agreement with previous both experimental and theoretical reports.\cite{hu2017probing, laturia2018dielectric, ermolaev2020giant}
For instance, our result for \ce{MoS2}: ${\Delta}n\sim1.34$ at 1550 nm is in agreement with previously obtained experimental values of ${\Delta}n\sim1.4$ (at 1530 nm, extracted by scattering scanning near-field optical microscopy (s-SNOM) \cite{hu2017probing}) and ${\Delta n}\sim1.5$ (in the infrared, extracted by spectroscopic ellipsometry \cite{ermolaev2020giant}). 
However, due to the lack of reported birefringence data for a broader range of available \glspl{tmd}, it is difficult to perform a comprehensive comparison. Our study partially fills this gap and, thus, provides an important contribution to the database of \gls{tmd} optical constants, useful for the development of future all-\gls{tmd} nanophotonic applications.  
Additionally, a combination of high-index ($n\gtrsim4$), low-loss and large birefringence ($\Delta n\gtrsim 1.4$) in the near-infrared range makes multilayer \glspl{tmd} a promising material platform for exploring photonic surface waves e.g., \textit{Dyakonov} \cite{takayama2009observation,takayama2014lossless} and \textit{Zenneck} \cite{babicheva2018near} surface waves.

\begin{figure}
\includegraphics[width=60mm]{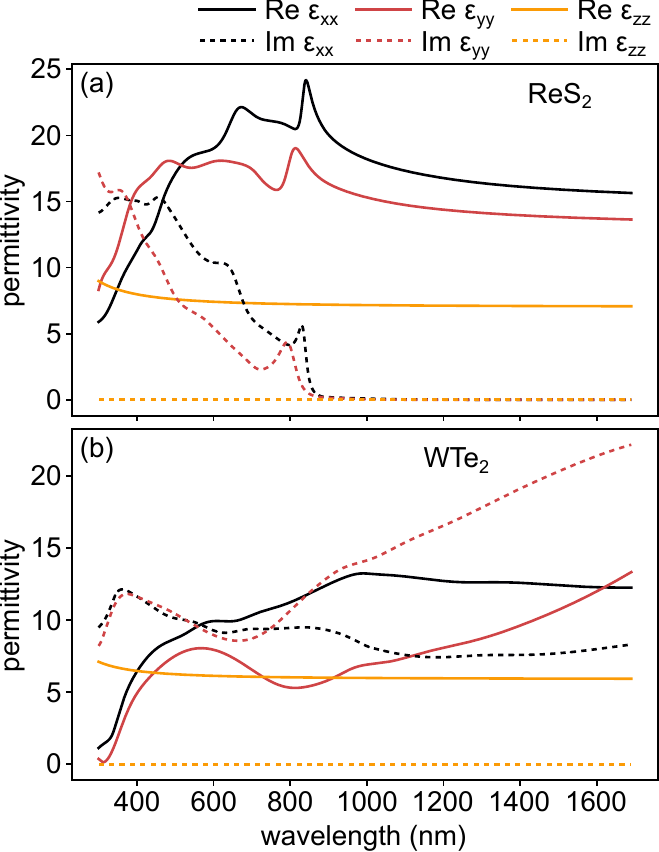}
\caption{Permittivity of bianisotropic materials: (a) semiconducting \ce{ReS2} and (b) lossy (metallic at low frequencies) \ce{WTe2} \gls{tmd} flakes. Corresponding exemplary Mueller Matrix measurements in \autoref{fig:mm-res2} and \autoref{fig:mm-wte2} show their biaxial nature.
}
\label{fig:bi}
\label{fig:semi-bi}
\label{fig:metal-bi}
\end{figure} 

\subsection{Biaxial semiconductor}

We now turn to the biaxial semiconductor -- \ce{ReS2}. The in-plane anisotropy of this material stems from the formation of covalent Re-Re bonds and correspondingly the 1T$'$-phase it adopts.\cite{aslan2016linearly}
This material has recently been theoretically predicted to have one of the highest permittivities in the visible -- near-infrared spectral range.\cite{shubnic2020high} Here, we report its experimentally measured optical constants.
Performing ellipsometry on such a material is more challenging than for uniaxial \glspl{tmd} from the measurement as well as analysis perspective. In-plane anisotropy requires rotation of the sample during the measurement to extract Euler angles of the material’s crystallographic structure. This, in turn, requires that the lateral size of a \ce{ReS2} flake should be larger than the beam spot for all in-plane directions. However, due to in-plane anisotropy \ce{ReS2} tends to shear-off along the $b$-axis \cite{wang2017cleavage} what makes exfoliation of a large symmetric flake extremely difficult. We were, however, able to prepare relatively large \ce{ReS2} samples with some folds which do not interfere with our measurements (\autoref{fig:flakes}). Moreover, interaction of polarized light with such a sample leads to a cross-polarization effect, thus $\Psi$ and $\Delta$ lose their meaning and the Generalized Ellipsometry, or more general Mueller Matrix ellipsometry, is required. To ensure sensitivity of the methods both to in-plane and out-of-plane components we performed a multisample analysis with thicknesses ranging from 200 nm to approximately 600 nm and at least two rotation angles with other measurement parameters being the same as for uniaxial semiconducting \glspl{tmd} above. It is worth mentioning that due to problems with exfoliation some of the \ce{ReS2} flakes are characterized by a terrace-like surface morphology and are not uniform in thickness. Thus in the analysis, data acquired for each rotation angle are treated as for a separate sample and its thickness (for every in-plane rotation angle) is a free fitting parameter. However, despite the presence of terraces, the crystalline axes of each flake are not disturbed and thus the fitted orientation (in-plane rotation) angles are consistent with the rotation angles during measurements. The in-plane permittivities $\varepsilon_{xx}$ and $\varepsilon_{yy}$ are described by multiple Tauc-Lorentz dispersion models, while the out-of-plane component is described by a single pole.
The experimentally obtained permittivity for \ce{ReS2} is shown in \autoref{fig:bi}a, while the fidelity of the fits is in \autoref{fig:fit-res2}.

Here we extracted two $n_{\parallel}$ and $n_{\perp}$ refractive indices of \ce{ReS2} at 1550 nm from the experimentally obtained permittivity, which yielded $n_{\parallel,xx}$: 3.97, $n_{\parallel,yy}$: 3.71, and $n_{\perp}$: 2.66, respectively (\autoref{fig:n}a,b).
\ce{ReS2} exhibits birefringence values of ${\Delta n_{xx} - n_{zz}}\sim1.31$ and ${\Delta n_{yy} - n_{zz}}\sim1.05$ (\autoref{fig:n}c). The obtained ${\Delta n_{yy} - n_{zz}}\sim1.05$ (\ce{ReS2}) is smaller than other TMDs. However, our data shows that \ce{ReS2} possess in-plane birefringence of ${\Delta n_{xx-yy}}$: $\sim$0.26 at 1550 nm, in addition to its out-of-plane anisotropy. This suggests that \ce{ReS2} could be an interesting material for next-generation photonics due to its both in-plane and out-of-plane anisotropic properties ($\mathrm{Re}(\varepsilon_{xx})\neq\mathrm{Re}(\varepsilon_{yy})\neq\mathrm{Re}(\varepsilon_{zz})\neq\mathrm{Re}(\varepsilon_{xx})$), together with high-index ($\geq 3.7$) and low-loss at telecom range. Moreover, a recent study\cite{kucukoz2022boosting} reports a light-induced phase transition in mono- and bilayers of \ce{ReS2}, which could be an additional benefit for \ce{ReS2}-based nanophotonic applications.

\subsection{Metallic and Hyperbolic (Uniaxial and Biaxial)}
The third group of samples investigated in the experiment are \glspl{tmd} that exhibit absorption in the whole measured wavelength range that comes from the metallic response and/or additional interband transitions. Due to a lack of sharp features in $\Psi$ and $\Delta$ spectra when placed directly on a reflective substrate, they require a special scheme of measurements to assure sufficient interaction of light with the samples and uniqueness of the ellipsometric models. This additional requirement is obtained by using a few-micron-thick thermally grown \ce{SiO2} layer on top of a Si substrate. When a thin semitransparent absorbing \gls{tmd} flake is deposited on such a support, interference in the \ce{SiO2} layer yields a distinct modulation of the ellipsometric signal, whose contrast depends on the thickness and extinction coefficient of the \gls{tmd} layer. This so-called interference approach was introduced by Hilfiker \emph{et al.} \cite{hilfiker2008survey} showing great improvement in the sensitivity of ellipsometric models applied to absorbing materials. 

In our experiment semitransparent flakes of thickness from $\sim$50 nm to $\sim$400 nm were exfoliated onto a 3 $\mu$m or 8 $\mu$m thick thermally grown \ce{SiO2} layer on a Si substrate. The measurements with the use of standard ellipsometry were carried out for the same wavelength and incidence angle range as for the previous samples (\emph{cf.} \autoref{fig:mm-nbse2}--\autoref{fig:mm-tase2} for Mueller matrix measurements confirming appropriate alignment of the uniaxial samples).
In the case of bianisotropic \ce{WTe2} the significant difference with respect to the details above was the need to perform in-plane rotation of the sample identically to \ce{ReS2}.
The anisotropic materials were modeled with Tauc-Lorentz and as needed Drude functions both for in-plane and out-of-plane components. The fitting procedure for these two types of materials is described in the Methods section. 

\begin{figure}
\includegraphics[width=60mm]{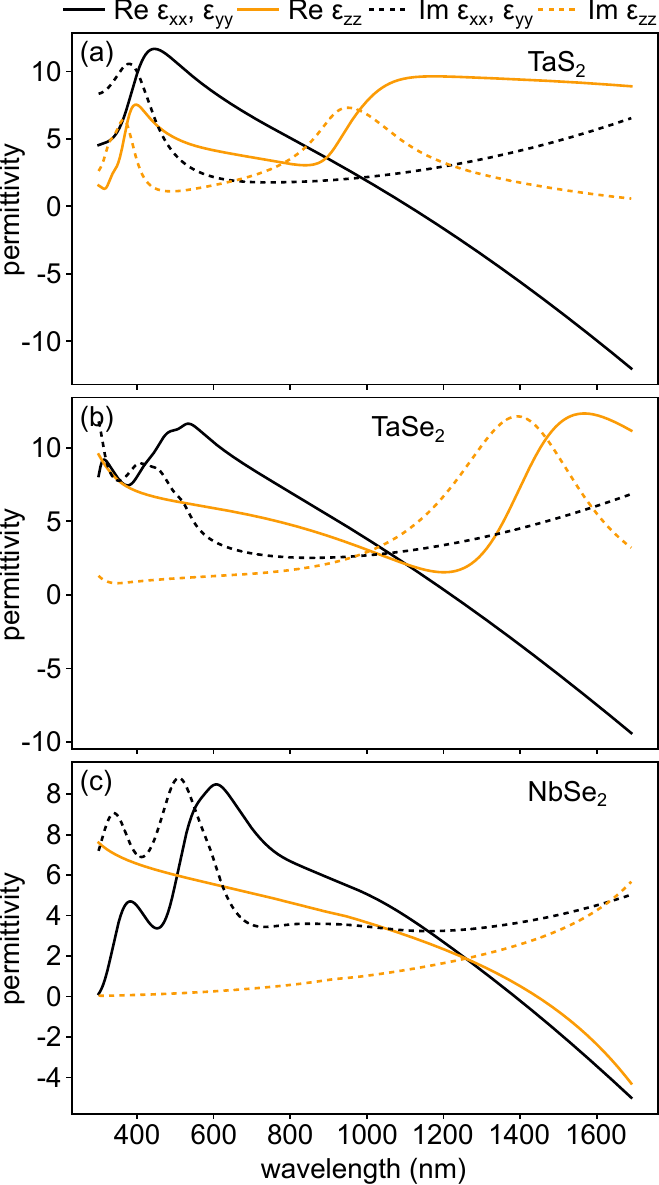}
\caption{Permittivity of uniaxial metallic \gls{tmd} flakes: (a) \ce{TaS2}, (b) \ce{TaSe2}, and (c) \ce{NbSe2}. See \autoref{fig:mm-nbse2}--\autoref{fig:mm-tase2} for Mueller Matrix measurements confirming their uniaxial properties and proper alignment.
}
\label{fig:metal-uni}
\end{figure} 

We now study uniaxial metallic \glspl{tmd}, e.g. \ce{TaS2}, \ce{TaSe2}, and \ce{NbSe2}, whose permittivities obtained in our measurements are shown in \autoref{fig:metal-uni}. Interestingly, the materials exhibit both dielectric and metallic responses in the studied spectral range, as indicated by a change of the sign of the values of the in-plane $\mathrm{Re}(\varepsilon)$ from positive to negative around 1100 nm to 1400 nm, depending on the material. Specifically, the plasma frequencies of the in-plane $\mathrm{Re}(\varepsilon)$ of Ta-based \glspl{tmd} \ce{TaS2} and \ce{TaSe2} are, respectively,  at $\sim$1110 nm ($\sim$1.11 eV) and $\sim$1217 nm ($\sim$1.01 eV), as illustrated in \autoref{fig:metal-uni}a,b. However, their out-of-plane $\mathrm{Re}(\varepsilon_{zz})$ remain positive showing only a dielectric response over the entire studied spectral range.

On the other hand, \autoref{fig:metal-uni}c shows that both in-plane and out-of-plane components of $\mathrm{Re}(\varepsilon)$ of \ce{NbSe2} are positive for wavelengths shorter than $\sim$1390~nm, suggesting that \ce{NbSe2} may behave like an out-of-plane anisotropic dielectric in the visible spectrum. Conversely, the real parts of the diagonal dielectric tensor become negative at longer wavelengths and retain the sign until the end of our measurement range. Furthermore, above $\sim$1200 nm \ce{NbSe2} is only weakly anisotropic.
The negative sign of the in-plane dielectric function is associated with a free-electron response just like for Ta-based metallic \glspl{tmd} and indicates an in-plane plasma frequency of $\sim$0.89~eV. However, the negative dielectric function in the out-of-plane direction is not associated with a free-electron, Drude-like response. Rather it is due to a high oscillator strength of a Lorentz-like response, which leads to the appearance of a Reststrahlen band for wavelengths above $\sim$1200 nm. The out-of-plane dielectric function should therefore become positive again at even longer wavelengths, however, this occurs outside of our measurement range ($>$1700 nm).

In contrast to the above three materials, \ce{WTe2} is an in-plane anisotropic \glspl{tmd} like \ce{ReS2}, but is metallic and hyperbolic. Recently, hyperbolic dispersion in \ce{WTe2} was demonstrated experimentally in the infrared region ($\sim$\unit[500]{cm$^{-1}$}) using far-field absorption measurements \cite{wang2020van}. Later, a new hyperbolic regime in the near-infrared ($\sim$1~eV) was theoretically predicted in monolayer \ce{WTe2} due to band-nested anisotropic interband transitions \cite{wang2020hyperbolicity}. However, such hyperbolicity of \ce{WTe2} in the near-infrared becomes weaker and disappears as the layer number increases from monolayer to bulk. Indeed, \autoref{fig:bi}b shows the  experimentally here-obtained permittivity from multilayer \ce{WTe2}. 
As expected, the real part of permittivity $\mathrm{Re}(\varepsilon)$ shows no sign of hyperbolic behavior in the studied spectral range.
However, we observed an interesting anisotropic behavior in the narrow spectral region around 800 nm, where $\mathrm{Re}(\varepsilon_{xx}) > \mathrm{Re}(\varepsilon_{zz}) > \mathrm{Re}(\varepsilon_{yy})$. 
This is not a common behavior in other \glspl{tmd}. Moreover, \ce{WTe2} possesses relatively large in-plane birefringence ranging from ${\Delta n_{x-y}}$: $\sim$0.2 (at 510 nm) to $\sim$1.12 (at 880 nm). The maximum obtained ${\Delta n_{x-y}}\sim$1.12 (\ce{WTe2}) is larger than a previously observed value of in-plane birefringence in \ce{ReS2} ($\sim$0.26).
Such large in-plane anisotropy may open new possibilities for nanophotonics, even though the imaginary part $\mathrm{Im}(\varepsilon_{xx})$ is relatively large.

\section{Conclusion}
In conclusion, we have experimentally measured both in-plane and out-of-plane optical constants from mechanically exfoliated \glspl{tmd} multilayers using spectroscopic ellipsometry over a broad spectral range of 300 nm -- 1700 nm. Our measurements include several semiconducting \ce{WS2}, \ce{WSe2}, \ce{MoS2}, \ce{MoSe2}, \ce{MoTe2}, as well as, in-plane anisotropic \ce{ReS2}, \ce{WTe2}, and metallic \ce{TaS2}, \ce{TaSe2}, and \ce{NbSe2} materials. The extracted parameters demonstrate a combination of several remarkable optical properties, such as a high-index ($n$ up till $\sim$4.84 for \ce{MoTe2}), significant anisotropy ($n_{\parallel}-n_{\perp} \approx 1.54$ for \ce{MoTe2}), and low absorption in the near infrared region. Moreover, metallic \glspl{tmd} show potential for a combined plasmonic-dielectric behavior and hyperbolicity, as their plasma frequencies occur in the $\sim$1000 -- 1300 nm range depending on the material. The knowledge of dispersive and anisotropic optical constants of these vdW materials opens new possibilities for future development of all-\glspl{tmd} nanophotonics.  

\section{Methods}

\noindent \textbf{Sample preparation:}
All \gls{tmd} flakes, including semiconducting and uniaxial \ce{WS2}, \ce{WSe2}, \ce{MoS2}, \ce{MoSe2}, (2H)\ce{MoTe2}, hyperbolic and metallic (2H)\ce{TaS2},  \ce{TaSe2}, \ce{NbSe2},  and biaxial \ce{ReS2} and \ce{WTe2}, were  mechanically exfoliated from bulk crystals (HQ-graphene) onto polydimethylsiloxane (PDMS) stamps using the scotch-tape method, and then transferred onto substrates using the all-dry transfer method \cite{Castellanos2014transfer, kinoshita2019dry}. For spectroscopic ellipsometry measurements, the lateral dimensions of the \gls{tmd} flakes should be larger than the beam size. To achieve proper sizes of \gls{tmd} flakes, we modified the previously developed method with few important concerns.

First, we chose the starting bulk crystals carefully, and exfoliate multilayers onto \gls{pdms} stamps only from large crystals (a centimeter at the least) with a homogeneous surface using a blue scotch-tape. Second, due to the thermoplastic properties of the \gls{pdms} film, adhesion between \gls{tmd} flakes and \gls{pdms} slightly decreases at the elevated temperature of 60~$^{\circ}$C. By exploiting this advantage, large multilayer \glspl{tmd} with relatively homogeneous thicknesses can be easily transferred onto a substrate for ellipsometric measurements. Thicknesses of the transferred \gls{tmd} flakes were measured using a VEECO profilometer, and we chose multilayer \glspl{tmd} with thicknesses ranging from a few tens of nanometers to microns. The minimum lateral size of uniaxial \gls{tmd} flakes investigated in this study is at least \unit[300]{$\mu$m} in one direction (beam width) and at least \unit[700]{$\mu$m} in the other to facilitate measurement at angles of incidence of up to (at the minimum of) 65~$^\circ$. In the case of biaxial \glspl{tmd}, \ce{ReS2} and \ce{WTe2}, both orthogonal in-plane directions need to be not smaller than \unit[400]{$\mu$m} to enable measurements up to (at least) 45~$^\circ$, larger if possible. For ellipsometric measurements of semiconducting \glspl{tmd}, one-side polished silicon substrates with a self-limiting natural oxide layer (1-3 nm) were used. In contrast to clear sharp excitonic features of the semiconducting, metallic TMDs lack such sharp features in $\Psi$ and $\Delta$ spectra. In order to perform high-quality measurements of metallic TMDs and to produce sharp interference features in the ellipsometric spectra, semitransparent metallic \gls{tmd} flakes of \ce{TaS2},  \ce{TaSe2},  \ce{NbSe2}, and \ce{WTe2} were transferred onto silicon substrates with a \unit[3]{$\mu$m} or \unit[8.8]{$\mu$m} thick thermally grown SiO$_2$ layer.

\noindent \textbf{Variable-angle spectroscopic ellipsometry:}
Ellipsometric measurements were carried out with the use of a Woollam RC2 dual rotating compensator (DRC) ellipsometer with a vertical auto angle stage. It allows for measurement of the full Mueller Matrix which is essential for analysis of bianisotropic samples and helpful for verifying that the uniaxial samples are correctly set up in the ellipsometer’s coordinates. Another feature that is accessible in the DRC architecture is the depolarization factor which allows for measurement and modeling of a sample's nonidealities like thickness nonuniformity or influence of a device's parameters/limitations like detector bandwidth or angular spread of the beam. Due to the small lateral dimensions of \gls{tmd} flakes they are measured with the use of focusing probes which reduce the light beam to a \unit[300]{$\mu$m} spot at normal incidence. The configuration of the ellipsometer with mounted focusing probes does not allow for transmission measurements, thus samples were prepared and measured in reflection up to a wavelength of \unit[1690]{nm} and the full accessible angle range from 20$^\circ$ to 75$^\circ$ depending of the size of the \gls{tmd} flake. Modeling and fitting were done with the use of CompleteEASE v6.61.

\noindent \textbf{Fitting of ellipsometric data:}
The investigated samples exhibit three types of optical responses which require different strategies in building appropriate models. The $\Psi$ curves of uniaxial semitransparent \gls{tmd} flakes exhibit interference maxima in the transparent regions. Proper fitting in these spectral ranges determines with very good accuracy both the thickness and the real part of the dielectric function, yielding a good starting point for further modelling of the data in the remaining spectral regions. For this type of samples we proceeded as follows:
\begin{itemize}
    \setlength\itemsep{-0.2ex}
    \item In the first step, an isotropic model is used and the transparent region is described by a Cauchy model $A+B/\lambda^{2}$, where the its parameters such as layer thickness, $A$ and $B$ are extracted by the Levenberg–Marquardt algorithm after fitting the model to the $\Psi$ and $\Delta$ curves.
    \item In the next step, the Cauchy model is converted to the Kramers-Kronig consistent B-spline curves with their subsequent expansion to the whole wavelength range. The B-spline function approximates the fitting curves with basic functions with their argument (photon energy, eV) equally spaced. We used a \unit[0.05]{eV} step in the whole energy range except where $\Psi$ exhibits sharp or anomalous behavior corresponding to e.g. exciton bands in the dielectric function. 
    \item In the third step, the isotropic model is converted to an anisotropic one and the out-of-plane component is described by a single UV pole. After minimization of the \gls{mse}, the B-spline model is parametrized by a General Oscillator model with the use of Tauc-Lorentz oscillators.
    \item The whole procedure is initially done for the thickest samples and after parametrization, the fitting procedure is repeated in a multisample analysis leading to a complex diagonal permittivity tensor which is common for all the samples. 
\end{itemize}
The lack of a transparent region for the metallic \gls{tmd} samples necessitates some changes to the above described procedure. 
First of all use of the interference approach requires a more complicated model taking into account the transparent interference layer as well as the additional interlayer present at Si and \ce{SiO2} interface. The Si substrate with thermal \ce{SiO2} was characterized prior to the final measurements and their extracted parameters were fixed in the initial stage of modeling. 
\begin{itemize}
    \setlength\itemsep{-0.2ex}
    \item The first step consists of fitting an isotropic Kramers-Kronig consistent Bi-spline function to the data for the thinnest samples exhibiting the most pronounced interference maxima. The thickness of the TMDs is fixed during the fitting procedure and their values are taken from profilometric measurements. 
    \item In the next step, the isotropic model is converted to an anisotropic one and the fitting procedure is repeated. For final improvement of the results roughness as well as sample and machine nonidealities are taken into account.
    \item In the last step the multisample analysis is carried out and the final Bi-spline model is parametrized with Tauc-Lorentz and Drude functions. 
\end{itemize}
   
Evaluation of the models is based on minimization of the \gls{mse}, the Correlation Matrix, and the Uniqueness test.
In case of bianisotropic samples the first and second procedures was used for, respectively, \ce{ReS2} and \ce{WTe2} with a modification that involved using a biaxial model instead of an uniaxial.

\section*{Associated Content}
\noindent
\textbf{Supporting Information}

\noindent
The Supporting Information is available free of charge at \url{https://pubs.acs.org/doi/10.1021/acsphotonics.}

\gls{si} Figures S1-S10 -- Exemplary Mueller Matrices for all materials; S11-S20 -- Agreement between measurements and fitted models. Tables S1 and S2 -- Summary of other fitting parameters; Note S1 -- Uncertainty of optical parameters accompanied by Figures S21-S23.\\

\section*{Funding}
B.M. and T.O.S. acknowledge financial support from the Swedish Research Council (under VR Milj\:o project, grant No: 2016-06059) and the Knut and Alice Wallenberg Foundation (grant No: 2019.0140). B.M. also acknowledges the European Research Council (ERC-CoG ``Unity'', Grant No: 865230). P.W. and T.J.A. thank the Polish National Science Center for support via the project 2019/34/E/ST3/00359 (T.J.A.)  and 2019/35/B/ST5/02477 (P.W.).

\section*{Notes}
\noindent
The authors declare no competing financial interest.

\noindent 
\textbf{Data availability:} The raw ellipsometric data along with fitted models in both the proprietary format of the CompleteEASE software (Woolam) and as plain text files as well as plain text files with fitted permittivities are available at \url{https://doi.org/10.5281/zenodo.6205431}.

\bibliography{Ellipsometry}

\end{document}


\begin{center}

{\large %
Supporting Information
}
\vspace{2ex}

{\Large\bf %
Optical constants of several multilayer transition metal dichalcogenides measured by spectroscopic ellipsometry in the 300-1700 nm range: high-index, anisotropy, and hyperbolicity
}
\vspace{3ex}

Battulga Munkhbat,$^{1,2,\ast}$ Piotr Wr\'obel,$^{3,\ast}$ Tomasz J. Antosiewicz,$^{3,1,\dagger}$ and Timur O. Shegai$^{1,\ddagger}$

\vspace{1ex}

\emph{
$^{1}$Department of Physics, Chalmers University of Technology, 412 96, Gothenburg, Sweden \\
$^{2}$Department of Photonics Engineering,
    Technical University of Denmark, 
    2800 Kongens Lyngby, Denmark \\
$^{3}$Faculty of Physics, University of Warsaw, Pasteura 5, 02-093 Warsaw, Poland \\
$^{\ast}$These authors contributed equally to this work. \\
$^{\dagger}$email: \href{mailto:tomasz.antosiewicz@fuw.edu.pl}{tomasz.antosiewicz@fuw.edu.pl} \\
$^{\ddagger}$email: \href{mailto:timurs@chalmers.se}{timurs@chalmers.se} \\
}

\end{center}

\vspace{10mm}

\tableofcontents

\clearpage

\section{\figurename{}s -- Examples of raw experimental data and ellipsometric fits}

\begin{figure}[h]
\includegraphics[width=1\textwidth]{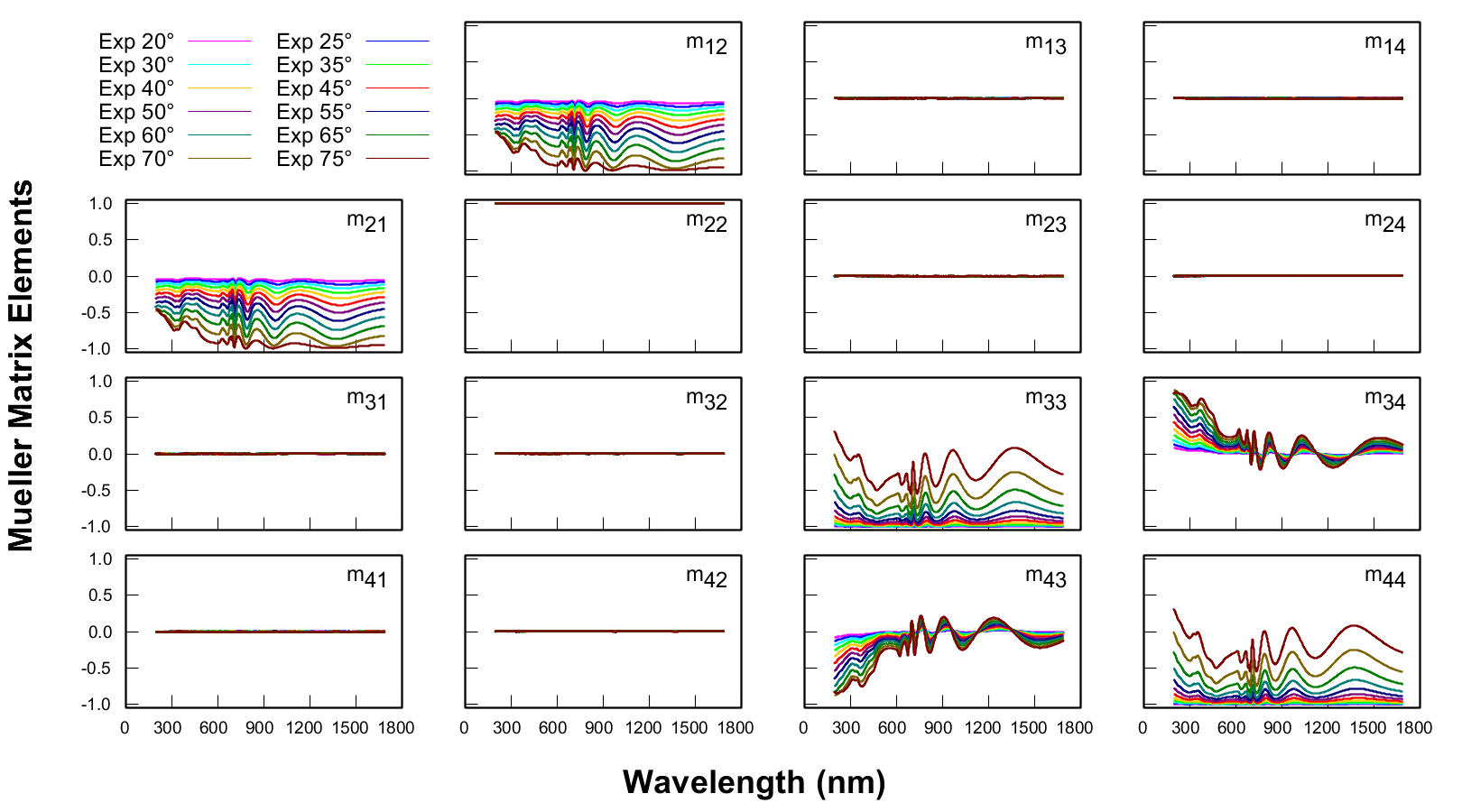}
\caption{Mueller Matrix elements for a selected \ce{MoS2} flake on a silicon substrate. Note that the zeros in the off-diagonal $2\times2$ sub-blocks prove the uniaxial nature of this material.
}
\label{fig:mm-mos2}
\end{figure}

\begin{figure}[h]
\includegraphics[width=1\textwidth]{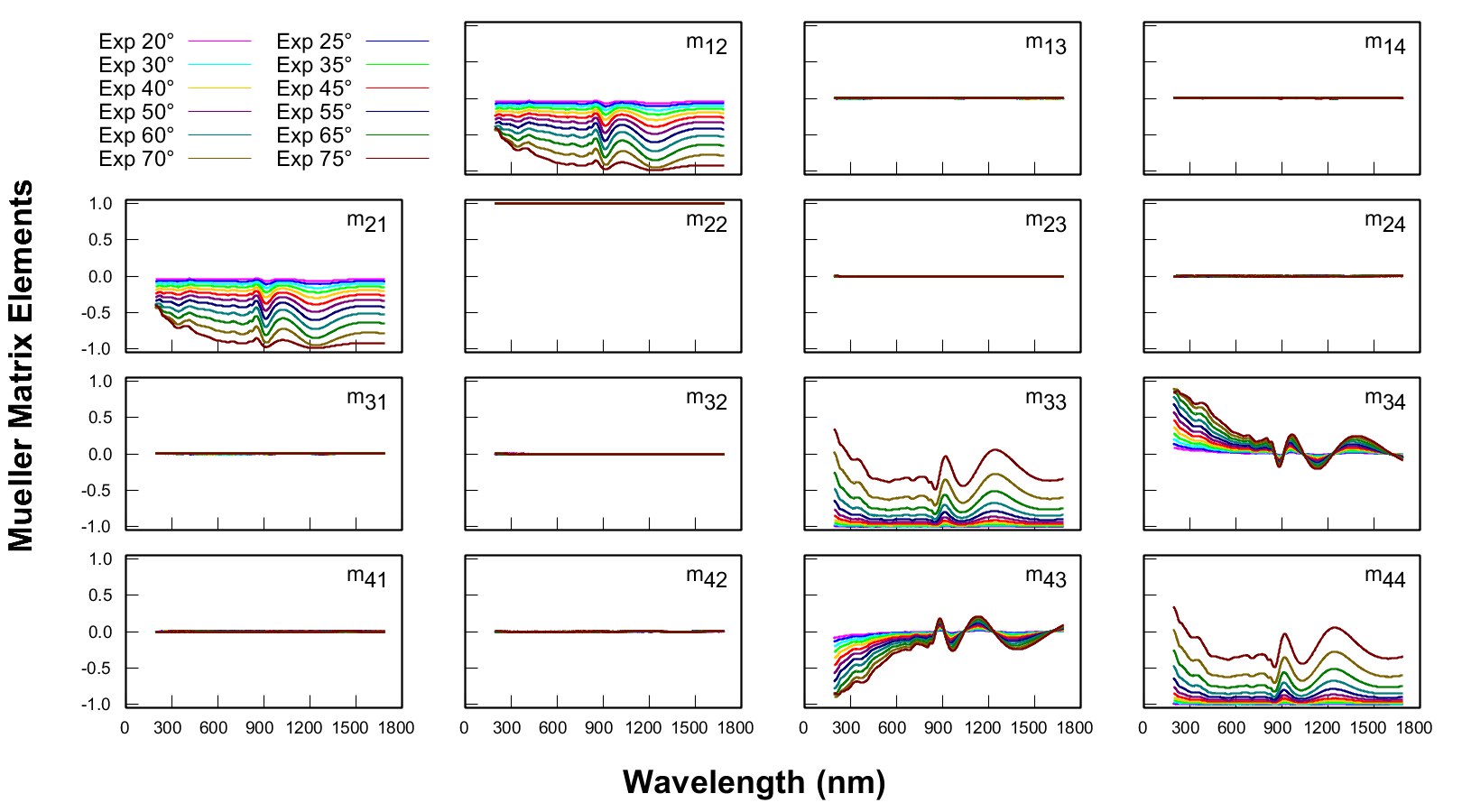}
\caption{Mueller Matrix elements for a selected \ce{MoSe2} flake on a silicon substrate. Note that the zeros in the off-diagonal $2\times2$ sub-blocks prove the uniaxial nature of this material.
}
\label{fig:mm-mose2}
\end{figure}


\begin{figure}[h]
\includegraphics[width=1\textwidth]{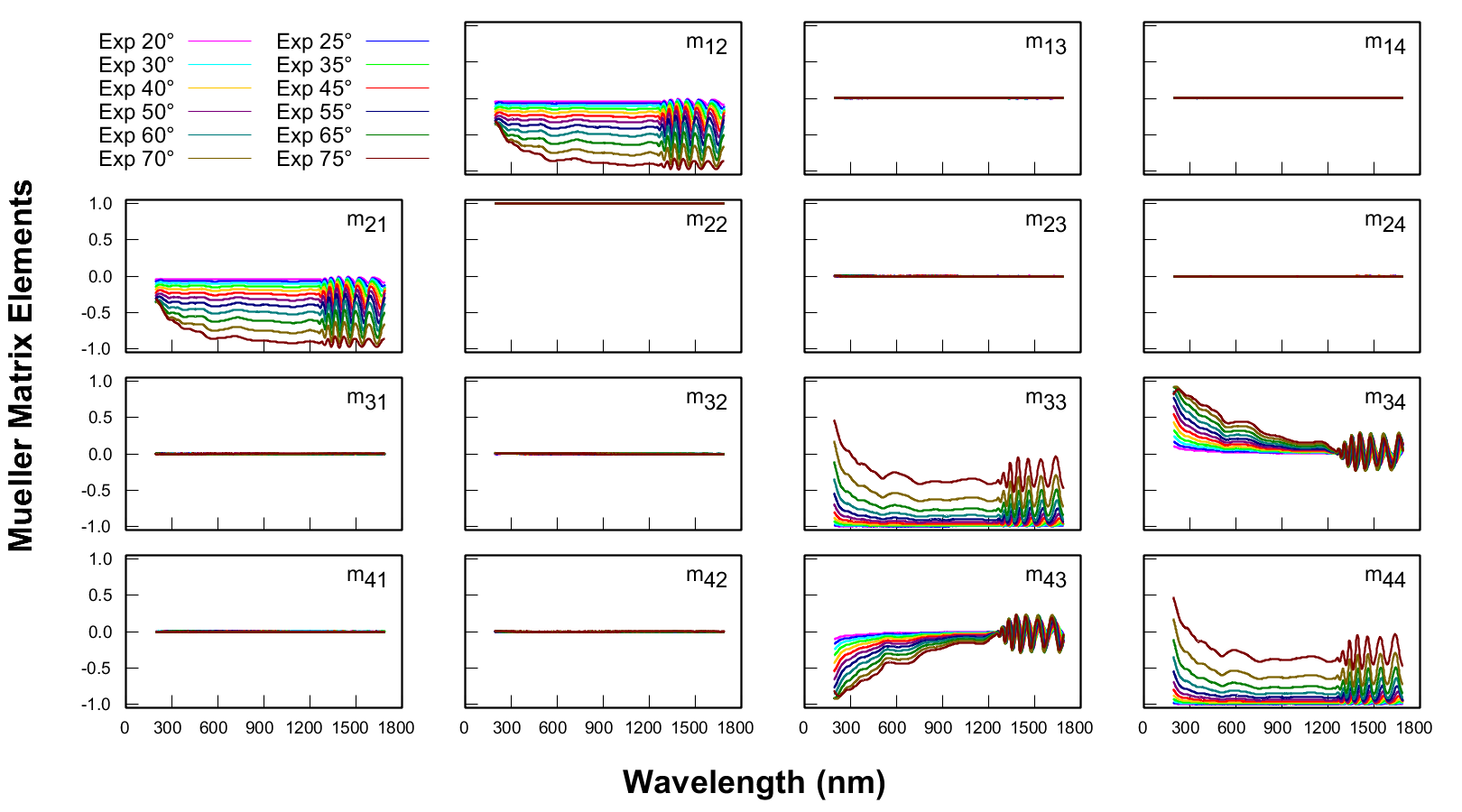}
\caption{Mueller Matrix elements for a selected \ce{MoTe2} flake on a silicon substrate. Note that the zeros in the off-diagonal $2\times2$ sub-blocks prove the uniaxial nature of this material.
}
\label{fig:mm-mote2}
\end{figure}

\begin{figure}[h]
\includegraphics[width=0.9\textwidth]{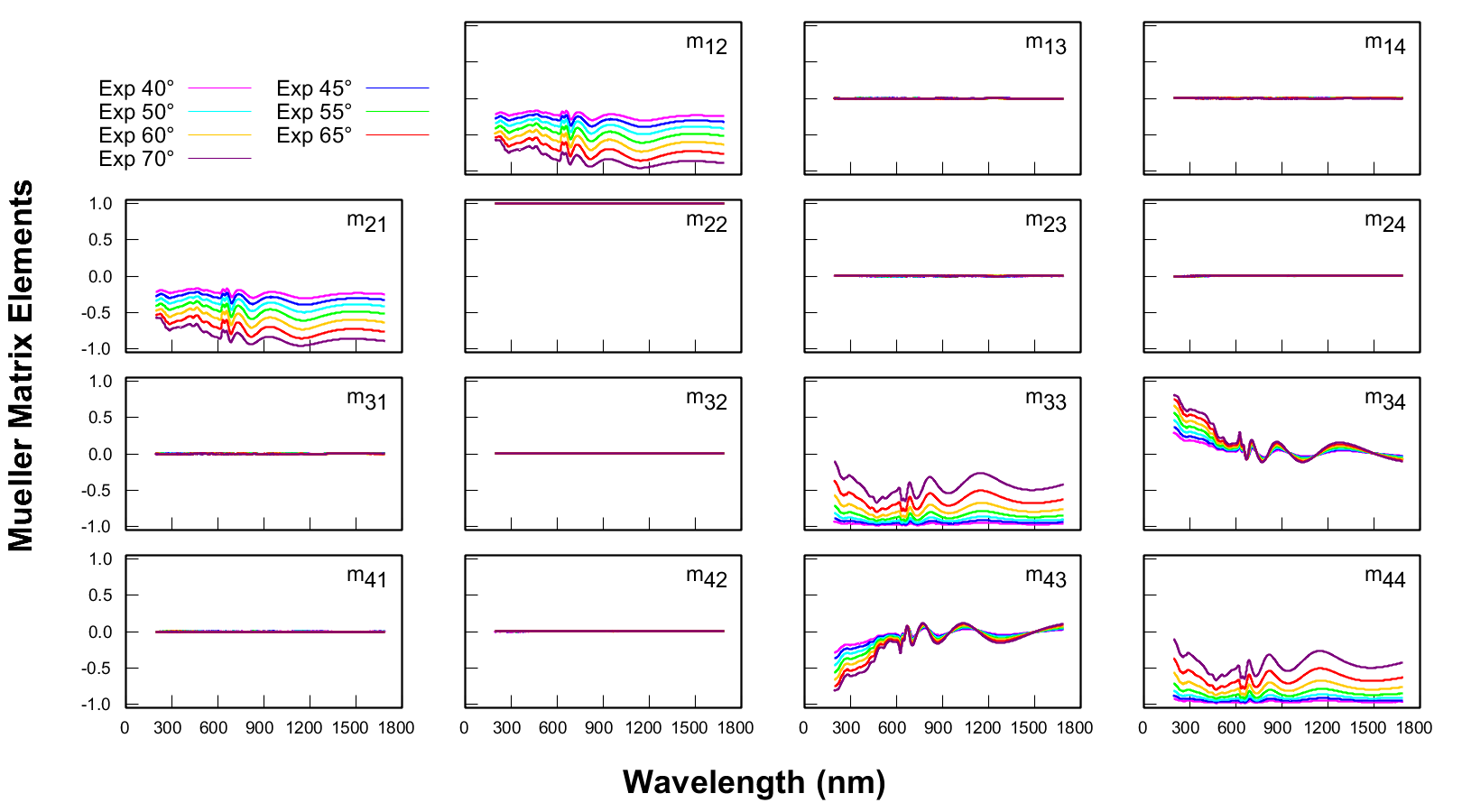}
\caption{Mueller Matrix elements for a selected \ce{WS2} flake on a silicon substrate. Note that the zeros in the off-diagonal $2\times2$ sub-blocks prove the uniaxial nature of this material.
}
\label{fig:mm-ws2}
\end{figure} 

\begin{figure}[h]
\includegraphics[width=1\textwidth]{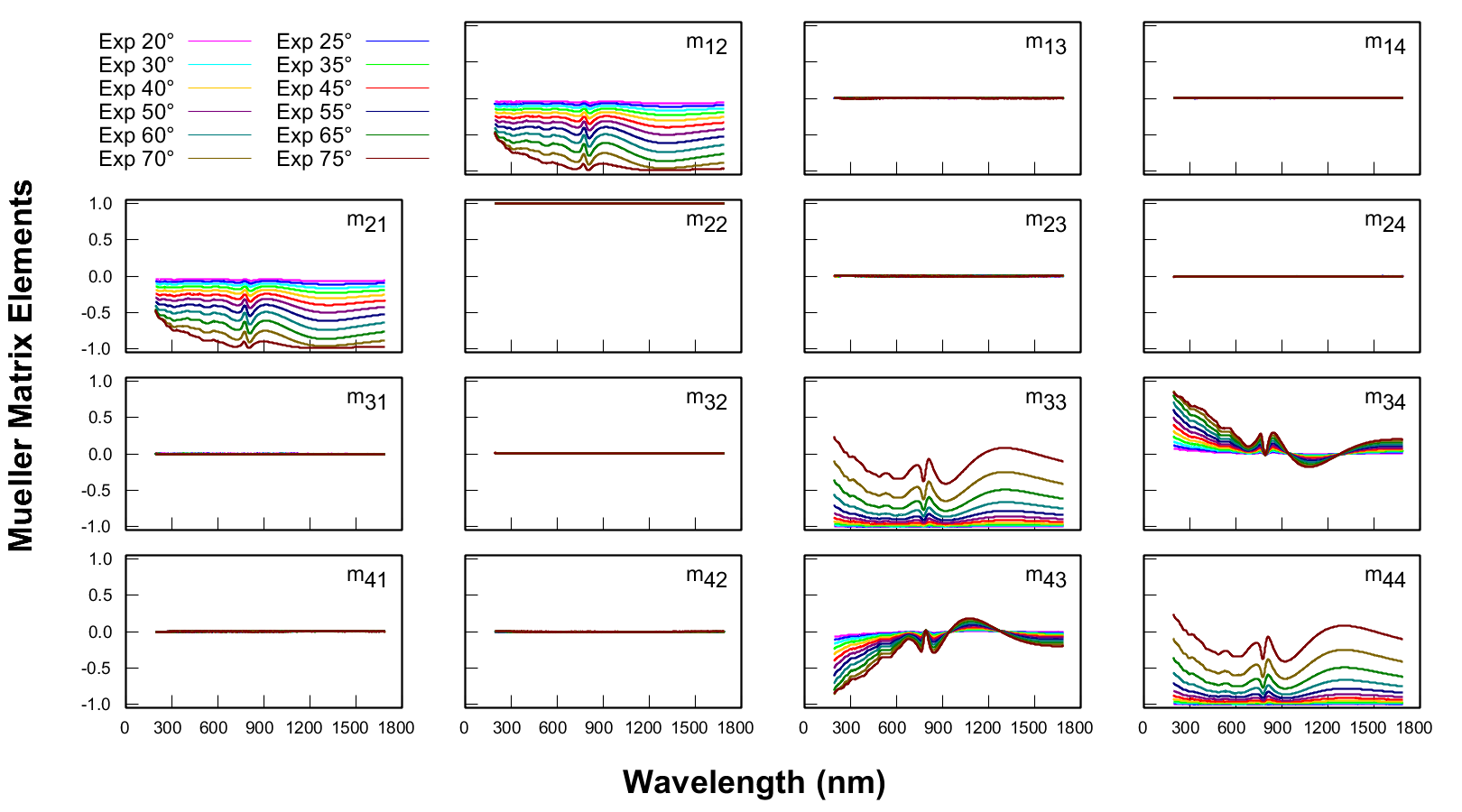}
\caption{Mueller Matrix elements for a selected \ce{WSe2} flake on a silicon substrate. Note that the zeros in the off-diagonal $2\times2$ sub-blocks prove the uniaxial nature of this material.
}
\label{fig:mm-wse2}
\end{figure} 

\begin{figure}[h]
\includegraphics[width=1\textwidth]{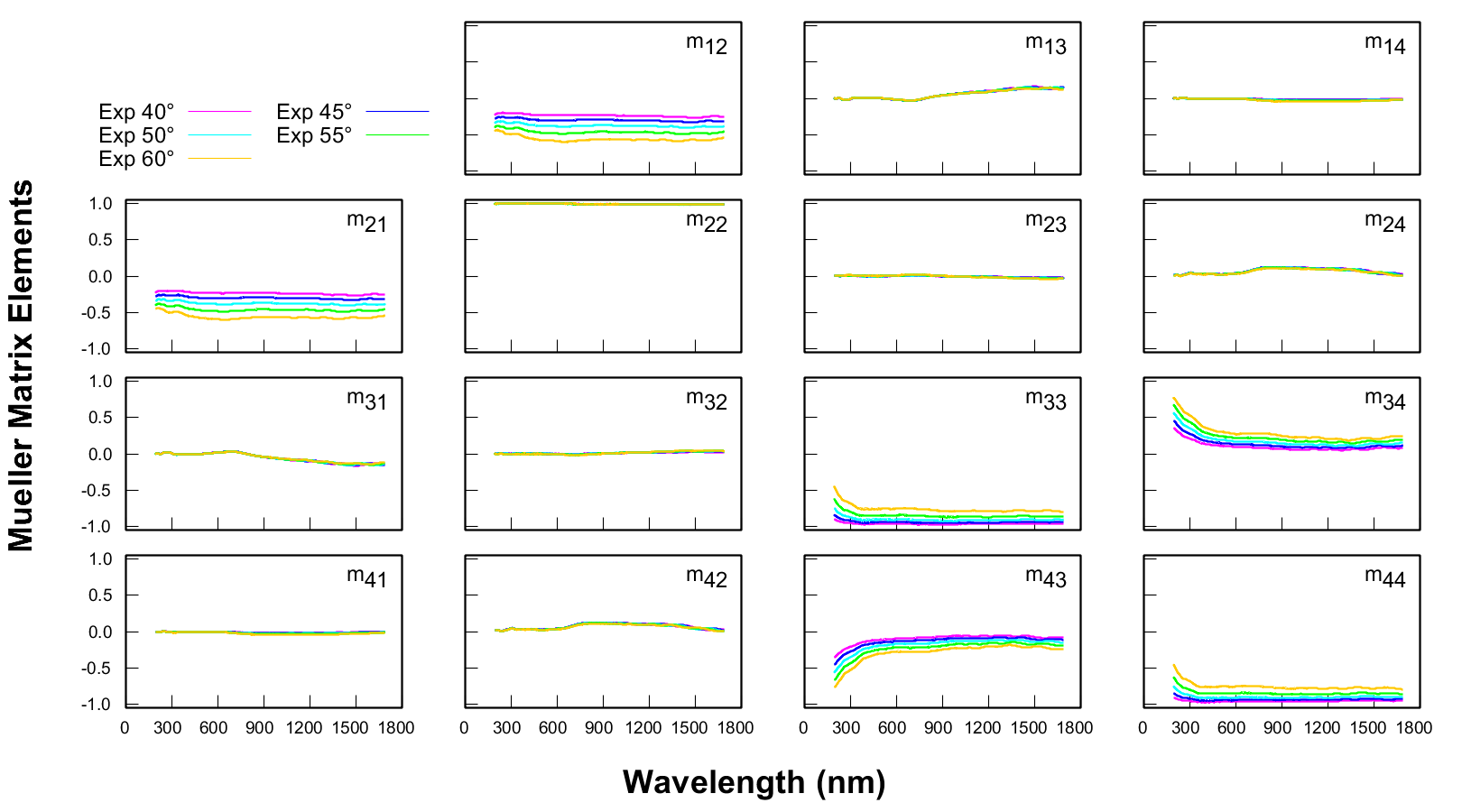}
\caption{Mueller Matrix elements for a selected \ce{WTe2} flake on a silicon substrate. Note that the off-diagonal $2\times2$ sub-blocks show the bianisotropic nature of this material.
}
\label{fig:mm-wte2}
\end{figure} 

\begin{figure}[h]
\includegraphics[width=1\textwidth]{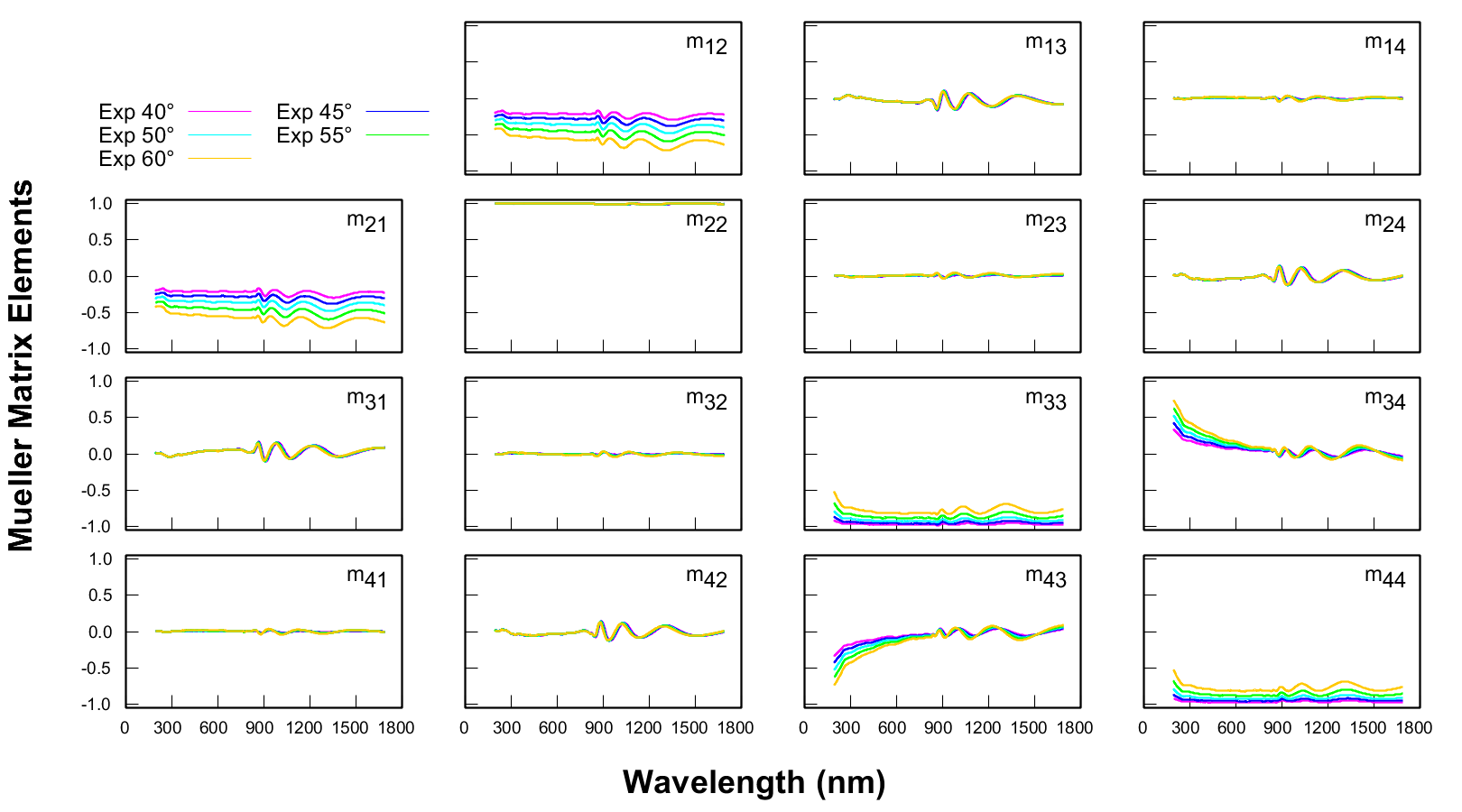}
\caption{Mueller Matrix elements for a selected \ce{ReS2} flake on a silicon substrate. Note that the off-diagonal $2\times2$ sub-blocks show the bianisotropic nature of this material.
}
\label{fig:mm-res2}
\end{figure} 

\begin{figure}[h]
\includegraphics[width=1\textwidth]{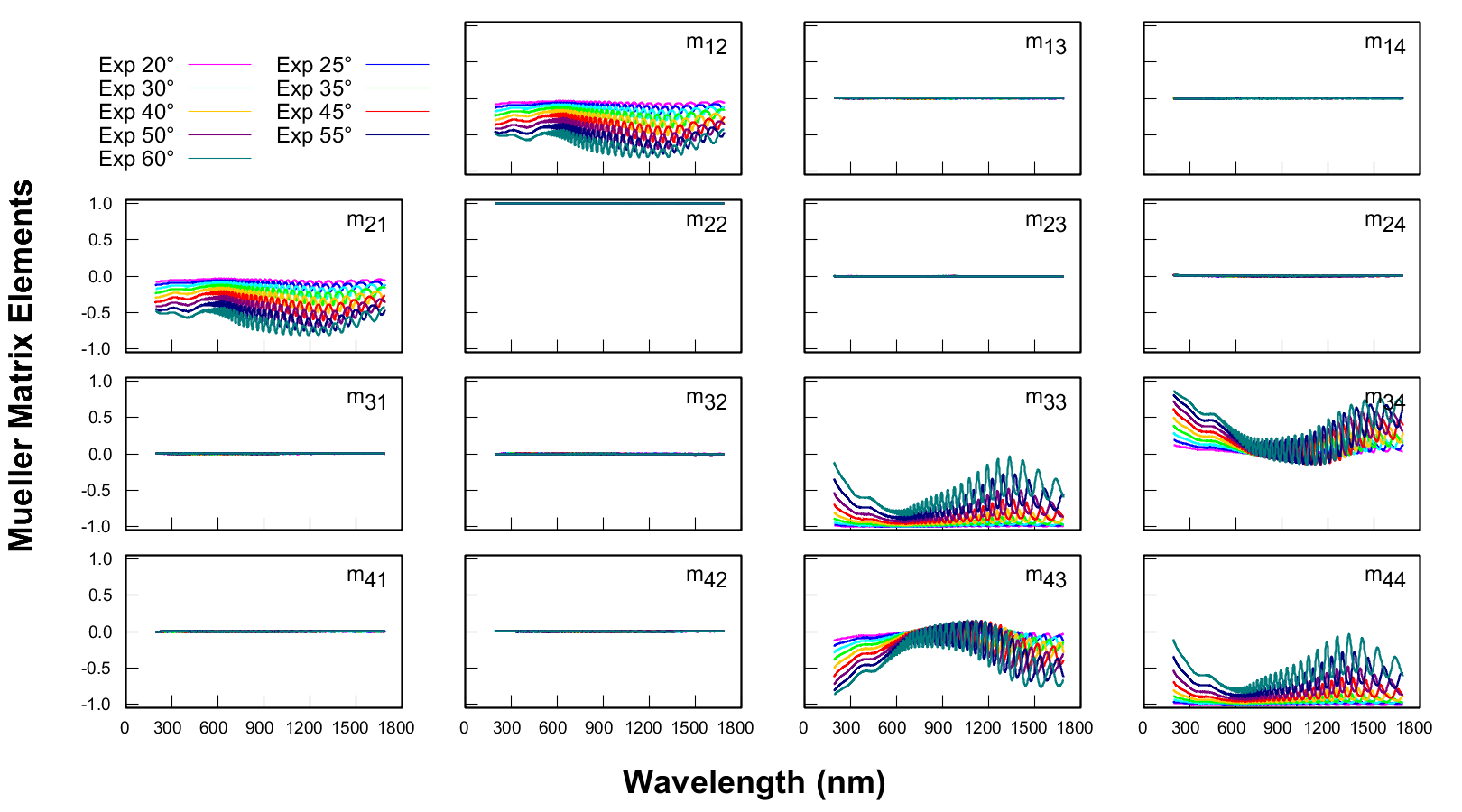}
\caption{Mueller Matrix elements for a selected \ce{NbSe2} flake on a silicon substrate. Note that the zeros in the off-diagonal $2\times2$ sub-blocks prove the uniaxial nature of this material.
}
\label{fig:mm-nbse2}
\end{figure} 

\begin{figure}[h]
\includegraphics[width=1\textwidth]{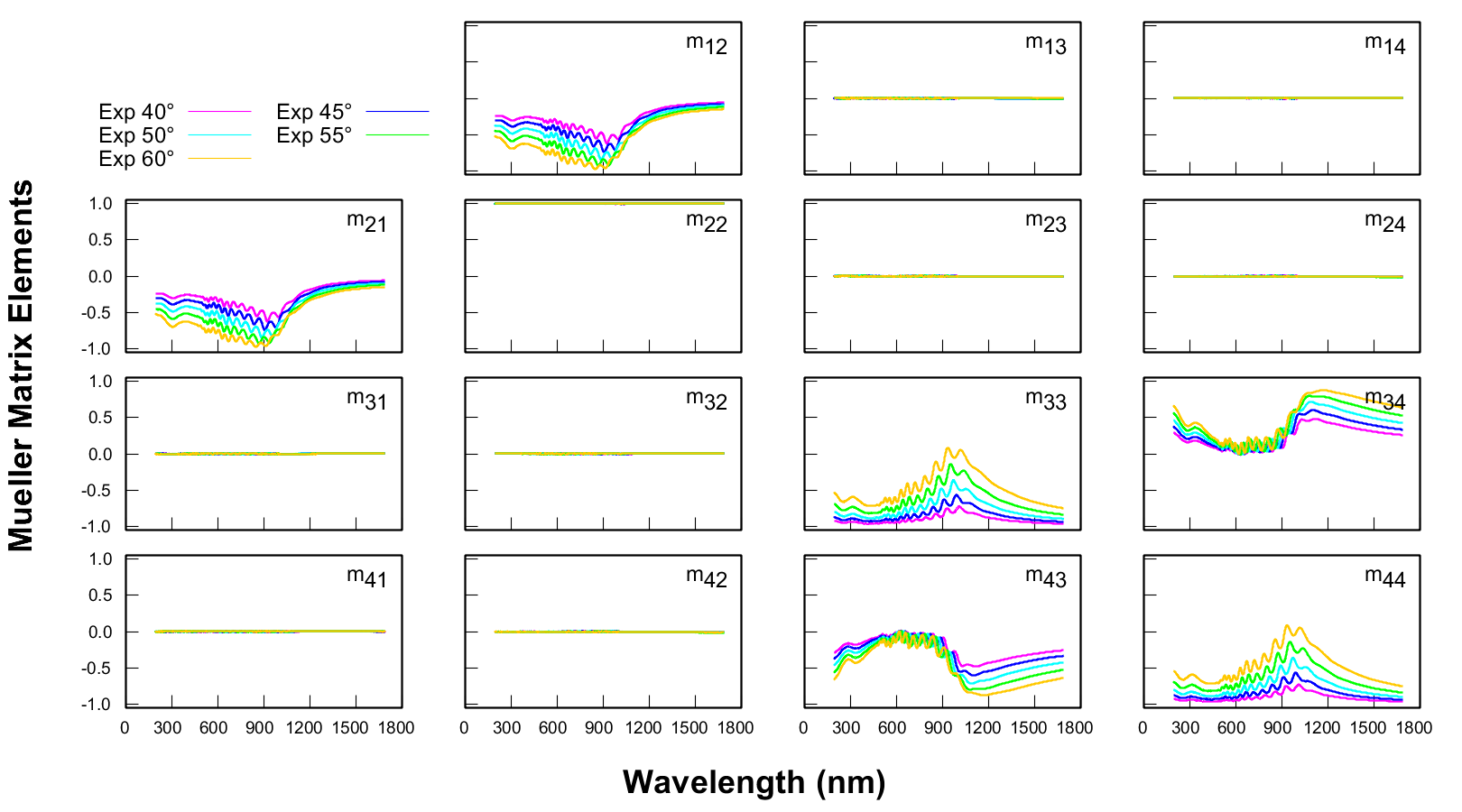}
\caption{Mueller Matrix elements for a selected \ce{TaS2} flake on a silicon substrate. Note that the zeros in the off-diagonal $2\times2$ sub-blocks prove the uniaxial nature of this material.
}
\label{fig:mm-tas2}
\end{figure} 

\begin{figure}[h]
\includegraphics[width=1\textwidth]{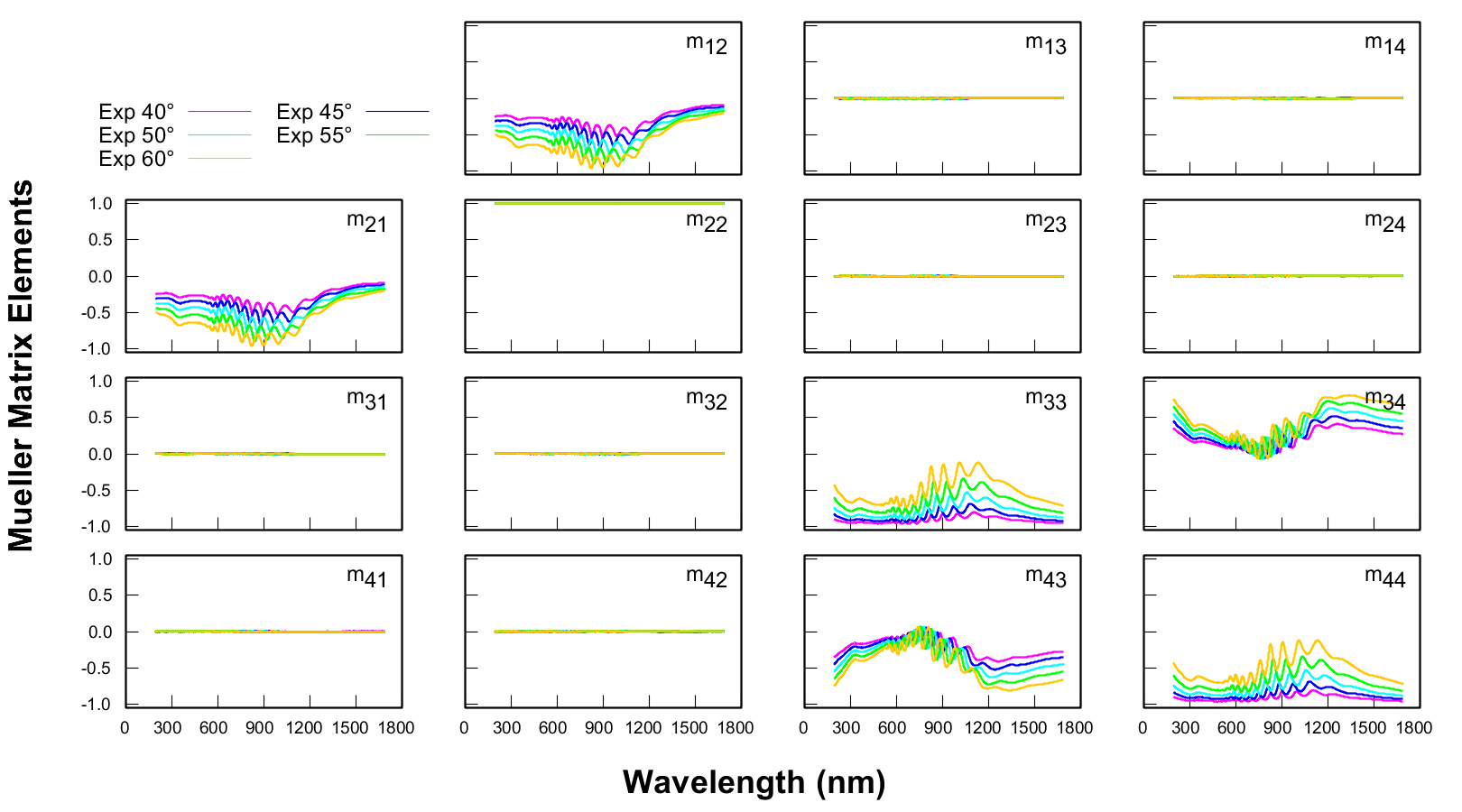}
\caption{Mueller Matrix elements for a selected \ce{TaSe2} flake on a silicon substrate. Note that the zeros in the off-diagonal $2\times2$ sub-blocks prove the uniaxial nature of this material.
}
\label{fig:mm-tase2}
\end{figure} 

\begin{figure}[h]
\includegraphics[width=1\textwidth]{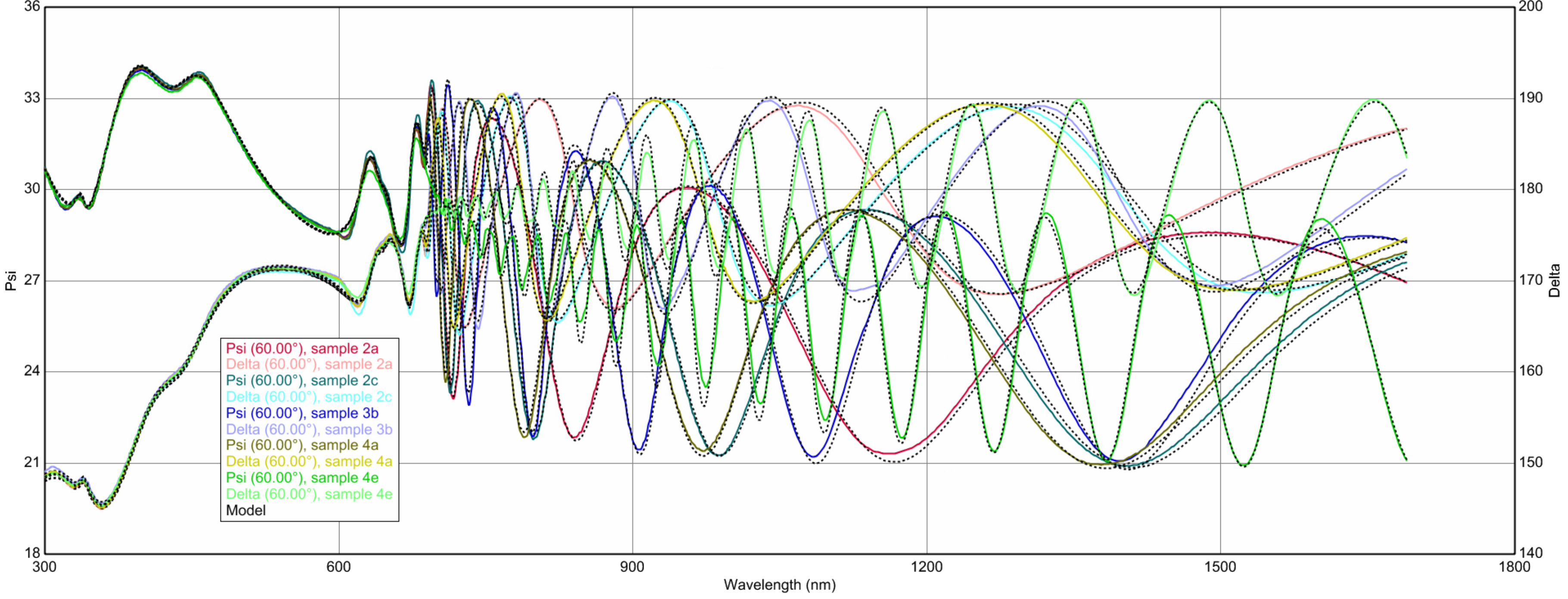}
\caption{Fidelity of the model fit to measurement data for \ce{MoS2}.
}
\label{fig:fit-mos2}
\end{figure} 

\begin{figure}[h]
\includegraphics[width=1\textwidth]{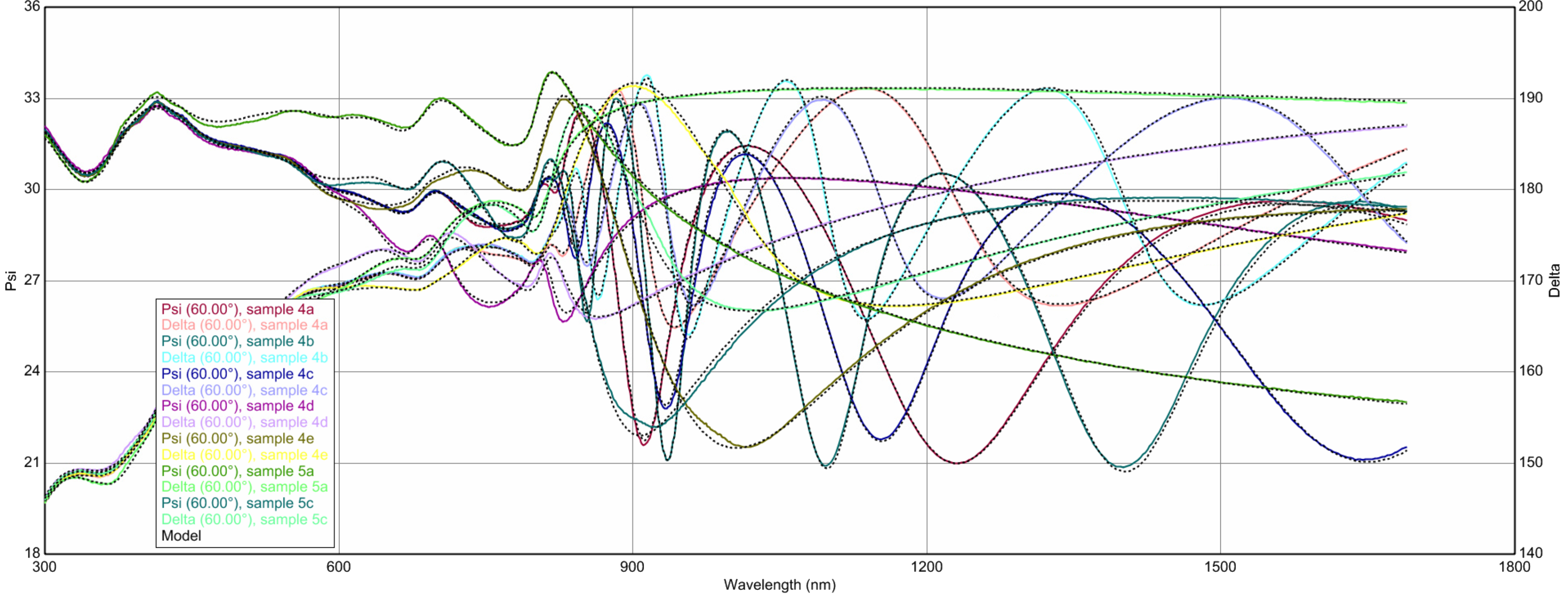}
\caption{Fidelity of the model fit to measurement data for \ce{MoSe2}.
}
\label{fig:fit-mose2}
\end{figure} 

\begin{figure}[h]
\includegraphics[width=1\textwidth]{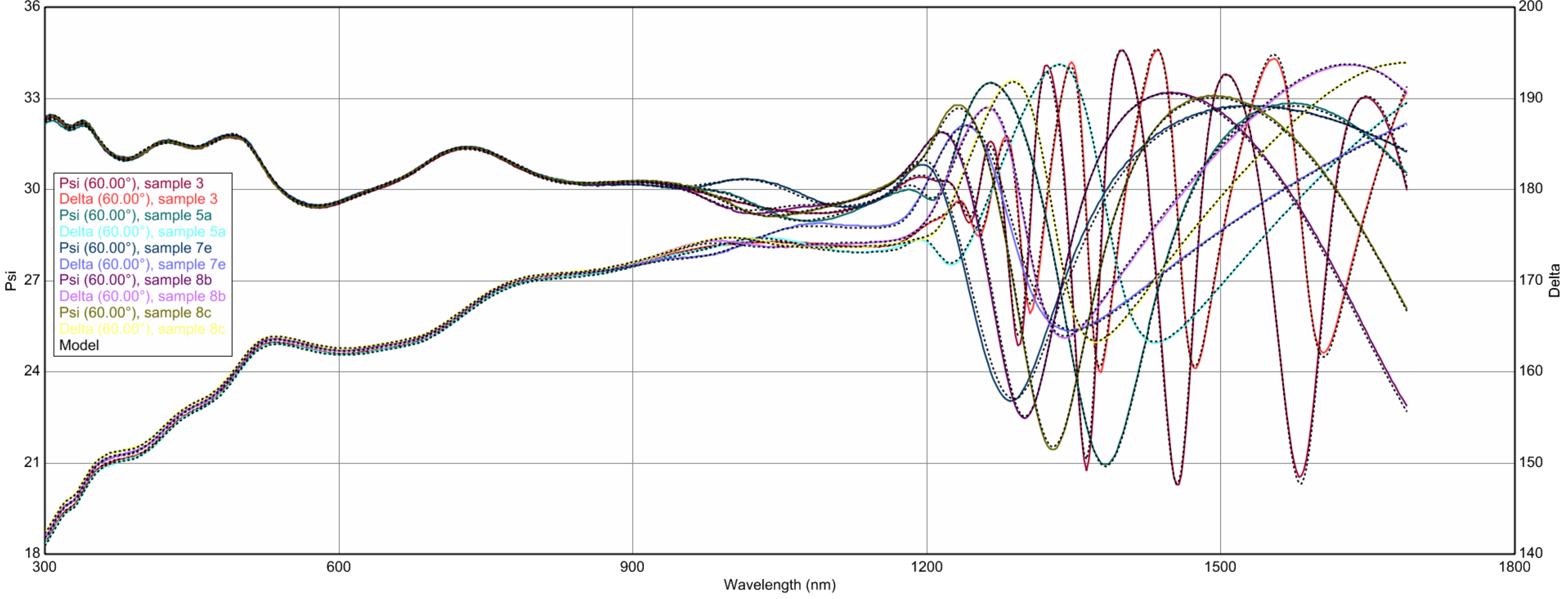}
\caption{Fidelity of the model fit to measurement data for \ce{MoTe2}.
}
\label{fig:fit-mote2}
\end{figure} 

\begin{figure}[h]
\includegraphics[width=1\textwidth]{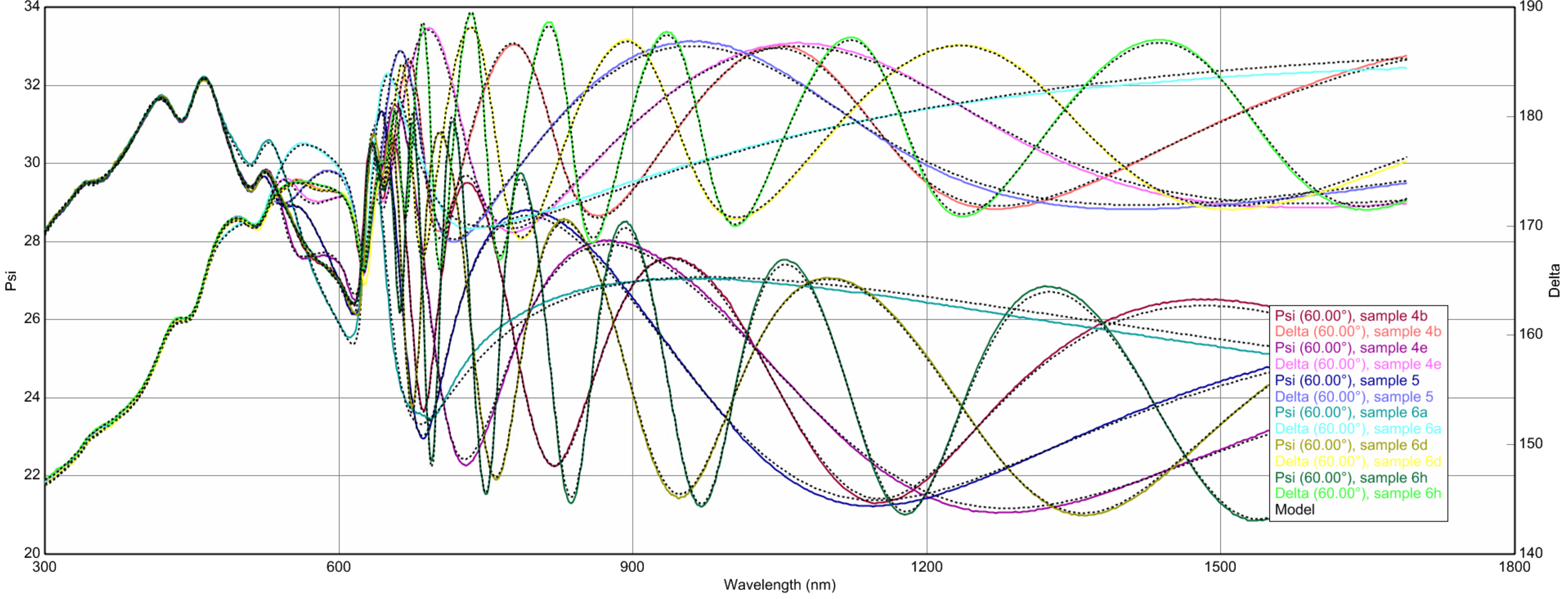}
\caption{Fidelity of the model fit to measurement data for \ce{WS2}.
}
\label{fig:fit-ws2}
\end{figure} 

\begin{figure}[h]
\includegraphics[width=1\textwidth]{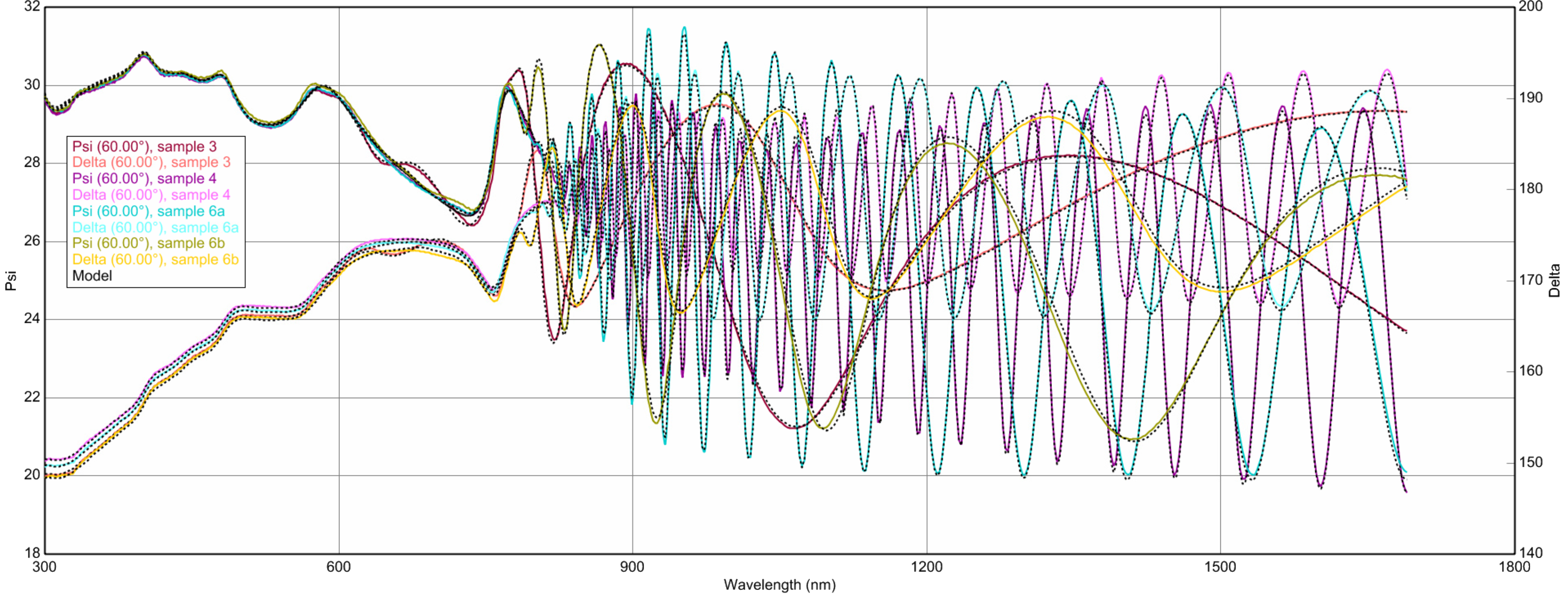}
\caption{Fidelity of the model fit to measurement data for \ce{WSe2}.
}
\label{fig:fit-wse2}
\end{figure} 

\begin{figure}[h]
\includegraphics[width=1\textwidth]{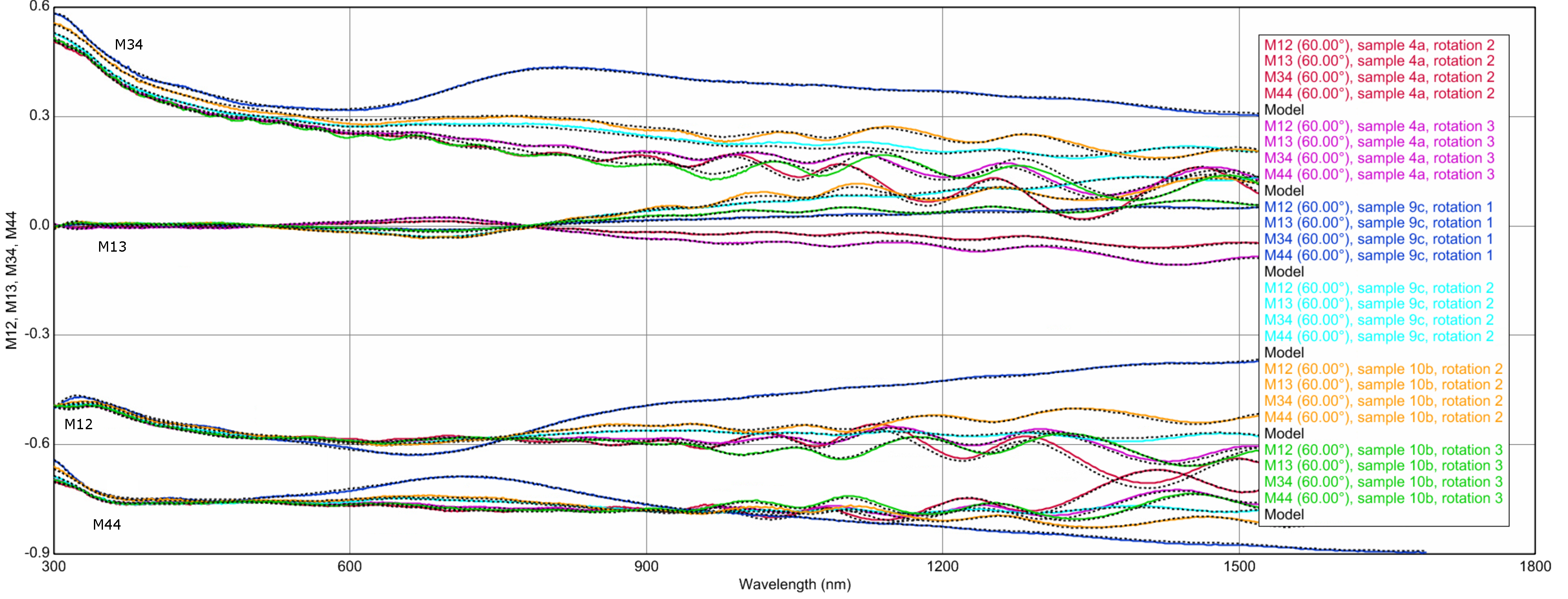}
\caption{Fidelity of the model fit to measurement data for \ce{WTe2}.
}
\label{fig:fit-wte2}
\end{figure} 

\begin{figure}[h]
\includegraphics[width=1\textwidth]{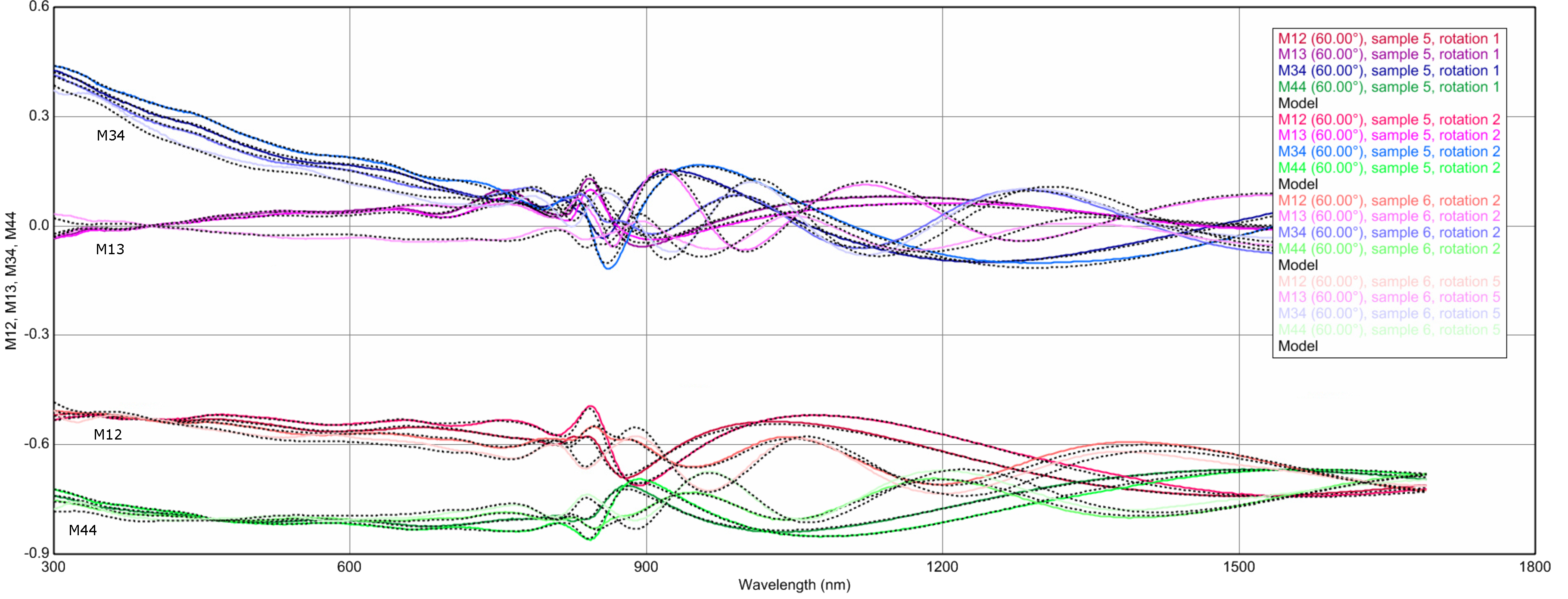}
\caption{Fidelity of the model fit to measurement data for \ce{ReS2}.
}
\label{fig:fit-res2}
\end{figure} 

\begin{figure}[h]
\includegraphics[width=1\textwidth]{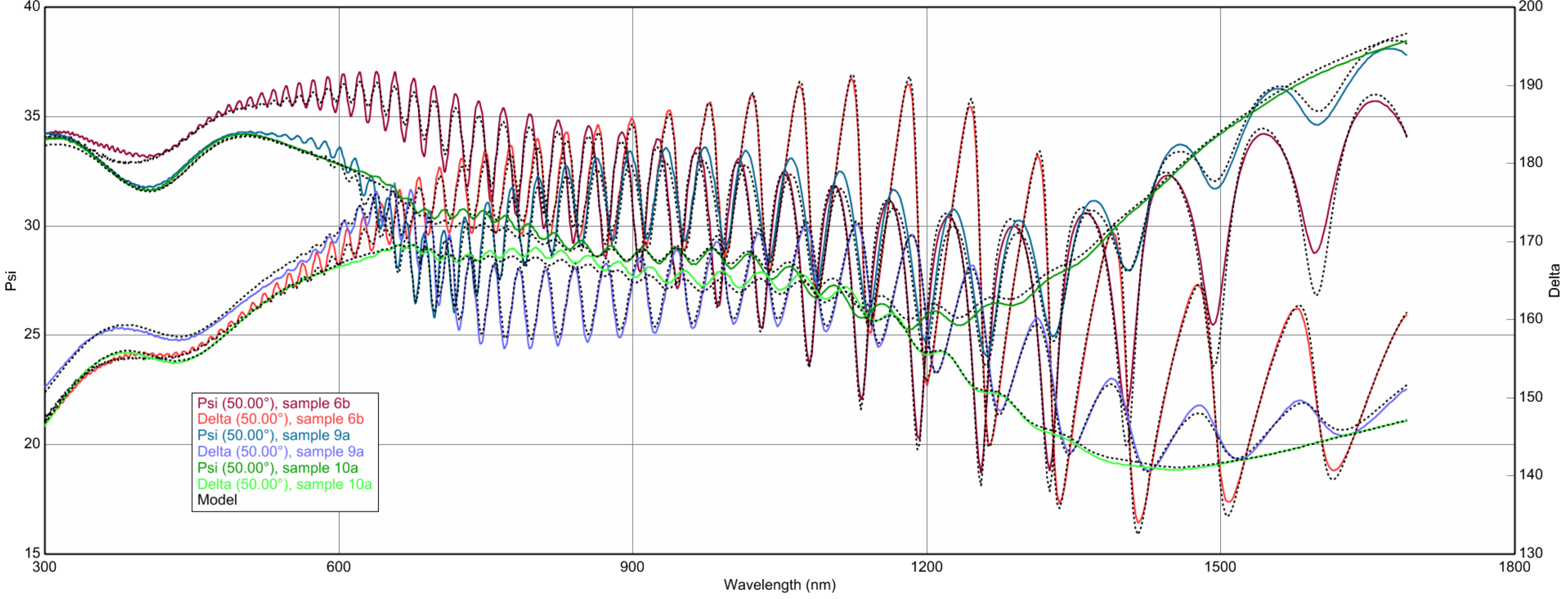}
\caption{Fidelity of the model fit to measurement data for \ce{NbSe2}.
}
\label{fig:fit-nbse2}
\end{figure} 

\begin{figure}[h]
\includegraphics[width=1\textwidth]{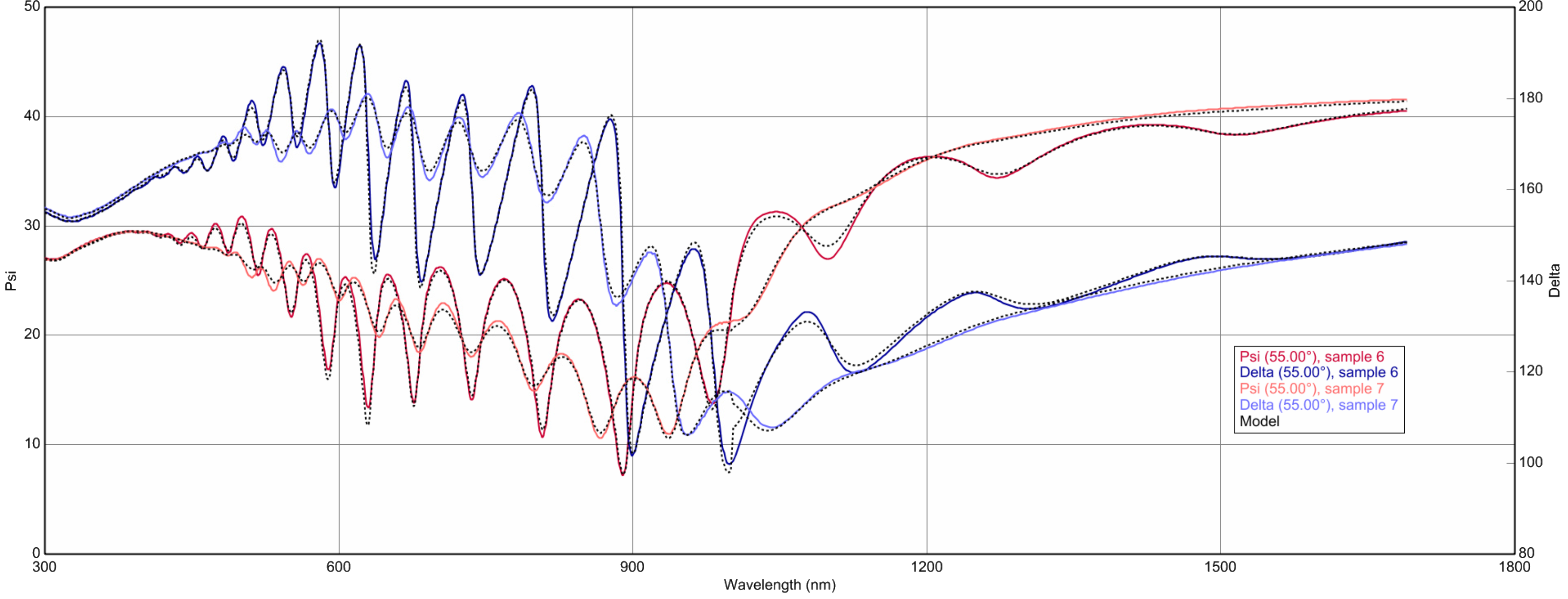}
\caption{Fidelity of the model fit to measurement data for \ce{TaS2}.
}
\label{fig:fit-tas2}
\end{figure} 

\begin{figure}[h]
\includegraphics[width=1\textwidth]{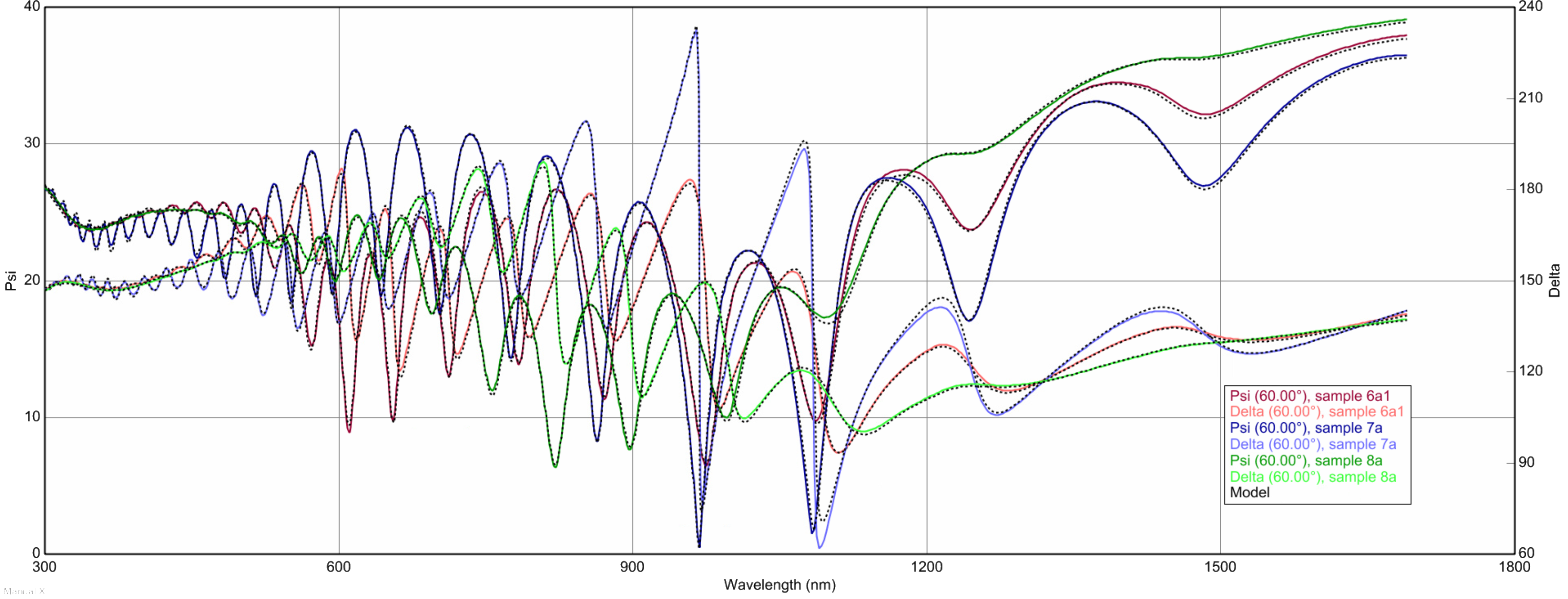}
\caption{Fidelity of the model fit to measurement data for \ce{TaSe2}.
}
\label{fig:fit-tase2}
\end{figure}

\clearpage
\section{Tables of ellipsometric models' parameters}

\begin{table}[H]
\centering
\titlecaptionlabel{Comparison of ellipsometric fitting parameters for uniaxial TMDs}{The fitted thickness values agree very well with (where available) reference thickness measurements obtained with a profilometer.}{tbl-1}
\begin{tabular}{ C{1.2cm} C{1.0cm} C{1.6cm} C{1.6cm} C{1.6cm} C{1.6cm}}
\toprule
 & & \multicolumn{3}{c}{Thickness} &  \\
\cline{3-5}
 &  & reference & fitted & nonunifor- & Roughness \\
Material & Sample & (nm) & (nm) & mity (\%) & (nm) \\
\toprule
\multirow{5}{*}{\ce{MoS2}} & 2a & \textcolor{black}{301} &  297 & 0 & 1.5 \\
 & 2c &  364 &  366 &   0 & 1.5 \\
 & 3b &  600 &  546 & 1.5 & 1.2 \\
 & 4a &  \textcolor{black}{359} &  359 &   0 & 1.3 \\
 & 4e & 1830 & 1778 & 3.4 & 1.2 \\
\hline
\multirow{10}{*}{\ce{MoSe2}} & 1 &  108 &  113 &   0 & 1.2 \\
 &  2 &  \textcolor{black}{123} &  124 &    0 & 0.8 \\
 & 4a &  324 &  301 &    0 & 1.2 \\
 & 4b &  538 &  524 &    0 & 1.1 \\
 & 4c &  \textcolor{black}{418}  &  418 &  7.6 & 1.1 \\
 & 4d &  \textcolor{black}{75}  &   76 & 22.4 & 0.8 \\
 & 4e &  118 &  118 &    0 & 1.3 \\
 & 5a &   24 &   30 & 20.2 & 0.9 \\
 & 5b &  \textcolor{black}{75}  &   77 & 23.6 & 2.0 \\
 & 5c &  \textcolor{black}{93}  &  100 &    0 & 1.1 \\
\hline
\multirow{6}{*}{\ce{MoTe2}} & 3 & 1600 & 1535 & 1.5 & 1.4 \\
 & 5a &  \textcolor{black}{428}  &  430 &    0 & 1.5 \\
 & 7e &  \textcolor{black}{259}  &  253 &  9.2 & 1.2 \\
 & 8b &  400 &  388 &  3.6 & 1.1 \\
 & 8c &  440 &  403 &  0.7 & 0.9 \\
\hline
\multirow{10}{*}{\ce{WS2}} & 3a & 1140 & 1165 &   0 & 1.2 \\
 & 3b &  \textcolor{black}{454}  &  456 &  0.1 & 1.4 \\
 & 3d &  \textcolor{black}{1356}  & 1356 &  0.1 & 1.3 \\
 & 4b &  \textcolor{black}{328}  &  325 &    0 & 1.4 \\
 & 4c &  \textcolor{black}{2325}  & 2247 &    0 & 1.3 \\
 & 4e &  \textcolor{black}{183}  &  184 &    0 & 1.5 \\
 &  5 &  164 &  163 &  0.1 & 1.5 \\
 & 6a &   91 &   81 &  0.1 & 1.5 \\
 & 6d &  \textcolor{black}{369}  &  389 &    0 & 1.5 \\
 & 6h &  \textcolor{black}{660}  &  664 &    0 & 1.4 \\
\hline
\multirow{4}{*}{\ce{WSe2}} & 3 & 300 & 273 & 4.5 & 1.5 \\
 &  4 &  \textcolor{black}{3601}  & 3829 &  0.3 & 0.8 \\
 & 6a &  --  & 2033 &  0.9 & 1.1 \\
 & 6b &  613 &  562 &  3.2 & 1.7 \\
\hline
\multirow{3}{*}{\ce{NbSe2}} & 6b & \textcolor{black}{52} & 52 & 0.4 & 6.1 \\
 & 9a &  \textcolor{black}{122}  &  122 &    0 & 3.5 \\
 &10a &  \textcolor{black}{373}  &  372 &    0 & 7.4 \\
\hline
\multirow{2}{*}{\ce{TaS2}} & 6 & 120 & 120 & 0.5 & 2.7 \\
 & 7 & 250 & 269 &    0 & 2.2 \\
\hline
\multirow{3}{*}{\ce{TaSe2}} & 6a1 & 102 & 102 & 0 & 3.3 \\
 & 7a &   74 &   70 &  0.7 & 3.8 \\
 & 8a &  \textcolor{black}{176}  &  173 &  0.2 & 3.6 \\
\bottomrule
\end{tabular}
\end{table}

\begin{table}[H]
\centering
\titlecaptionlabel{Comparison of ellipsometric fitting parameters for bianisotropic TMDs}{The fitted thickness values agree very well with (where available) reference thickness measurements obtained with a profilometer, as do the relative rotation angles. $^\dagger$Note, that sample 7 of \ce{ReS2} is L-shaped and the two arms have different thicknesses as confirmed by profilometer measurements and ellipsometry fitting.}{tbl-2}
\begin{tabular}{ C{1.1cm} C{1.0cm} C{1.6cm} C{1.6cm} C{1.6cm} C{1.2cm} C{1.5cm} C{1.5cm}}
\toprule
 & & \multicolumn{3}{c}{Thickness} & Rough- & \multicolumn{2}{c}{Orientation angle} \\
\cline{3-5}\cline{7-8}
 &  & reference & fitted & nonunifor- & ness & reference & fitted \\
Material & Sample & (nm) & (nm) & mity (\%) & (nm) & (degrees) & (degrees) \\
\toprule
\multirow{6}{*}{\ce{ReS2}} & \multirow{2}{*}{5} & \multirow{2}{*}{196} &  203 & 1.2 & 1.5 & -1.6 & 35.8 \\
 & & &  208 & 2.7 & 1.6 & -16.2 & 21.8 \\
\cmidrule{2-8}
 & \multirow{2}{*}{6} & \multirow{2}{*}{483} &  483 & 2.5 & 1.6 & 15.3 & 46.4 \\
 & & &  512 & 2.9 & 1.8 & 81.2 & 110.4 \\
\cmidrule{2-8}
 & \multirow{2}{*}{7$^\dagger$} & \textcolor{black}{191} &  190 & 0.5 & 1.4 & -22.3 & -7.7 \\
 & & \textcolor{black}{681} &  645 & 5.9 & 1.8 & 42.3 & 57.7 \\
\hline
\multirow{11}{*}{\ce{WTe2}} & \multirow{3}{*}{4a} & \multirow{3}{*}{\textcolor{black}{256}} &  233 & 0.9 & 0.2 & 19.6 & 10.7 \\
 & & &  251 & 3.4 & 0.2 & 31.1 & 22.0 \\
 & & &  235 & 1.9 & 0.4 & 46.0 & 37.9 \\
\cmidrule{2-8}
 & \multirow{3}{*}{5} & \multirow{3}{*}{\textcolor{black}{410}} &  403 & 1.3 & 0.3 & -1.5 & -86.3 \\
 & & &  417 & 1.6 & 0.5 & 41.5 & -41.7 \\
 & & &  460 & 1.6 & 0.4 & 62.8 & -20.2 \\
\cmidrule{2-8}
 & \multirow{3}{*}{9c} & \multirow{3}{*}{\textcolor{black}{440}} &  397 & 1.1 & 0.4 & 0.8 & -80.4 \\
 & & &  412 & 1.9 & 0.3 & 46.3 & -35.0 \\
 & & &  372 & 6.6 & 0.3 & 86.6 & 6.1 \\
\cmidrule{2-8}
 & \multirow{2}{*}{10b} & \multirow{2}{*}{\textcolor{black}{233}} &  279 & 0.4 & 0.7 & 45.1 & -40.4 \\
 & & &  272 & 1.5 & 0.2 & 73.0 & -15.3 \\
\bottomrule
\end{tabular}
\end{table}

\FloatBarrier

\vspace{2cm}
\section{Supporting Notes}

\vspace{0.25cm}
\noindent
\notetitlelabel{note:uncertainty}{Uncertainty of optical constants}

Evaluation of inaccuracies of extracting the optical parameters from ellipsometric measurements is a very challenging task and cannot be done unequivocally, since the technique is not a direct method. The uncertainty can be influenced by systematic errors of the measurement itself like beam divergence, angle of incidence, sample position, etc, however, the biggest impact on the trustworthiness of the extracted data stems from the optical model. Its validity can be judged by the fit quality and statistically by the Mean Squared Error (MSE), which is a prime parameter used for evaluating the model and how it fits the measured data. However, the MSE parameter alone is merely an indicator that in the case of yielding small values the model might be true. More importantly, it does not take into account what is the sensitivity of the model to the various fitting parameters.  

Thus, another way to evaluate the model is the uniqueness test. In this test of the model a given parameter, for example the layer thickness, is fixed (in a given range) at various values while all the other ones are fitted. The width (or more genrally the shape) of the minimum of the MSE curve as a function of the chosen parameter shows the uncertainty of the model. However, considering the fact that many parameters of the model influence the final result and their impact may be vary depending on a particular spectral range, using this measure alone for the evaluation of uncertainties of optical constants may be insufficient as well.  

\begin{figure}[t]
\centering\includegraphics[width=0.99\columnwidth]{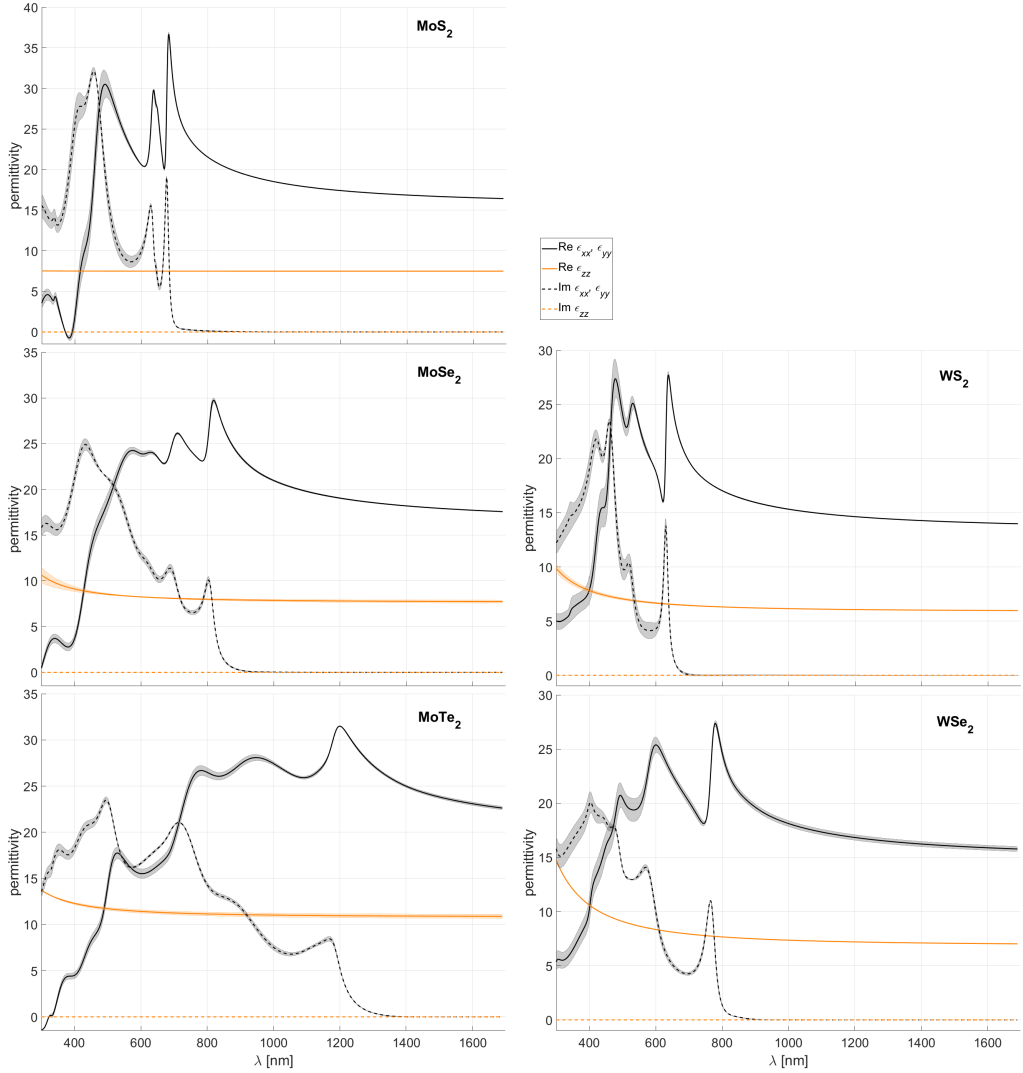}
\caption{Uncertainty of the complex permittivity for uniaxial semitransparent TMDCs.
}
\label{fig:Errors-uniaxial}
\end{figure}

A more systematic way of evaluating the goodness of the proposed models and their fits is done using a Fit Parameter Error Estimation tool that is included in the analysis software (CompleteEASE). 
(i) The first procedure is based on analyzing random errors to show the influence of the measurement uncertainties. It reanalyzes a randomized set of trial experimental data based on the original measurements and refits many sets with the same model. This gives an estimate of the uncertainty for individual fitting parameters as well as the optical constants. (ii) Another set of procedures evaluates the influence of systematic errors coming from such sources as angle offset, wavelength shift, $\Psi$, and $\Delta$ offset or inaccuracies of the complex refractive index of the substrate. (iii) Finally, magnitudes of the fitting errors are tested by adding or subtracting the same magnitude of error at every wavelength to the existing data set and refitting the manipulated data. 

\begin{figure}[t]
\begin{minipage}[b]{.49\textwidth}
\centering
\includegraphics[width=1\textwidth]{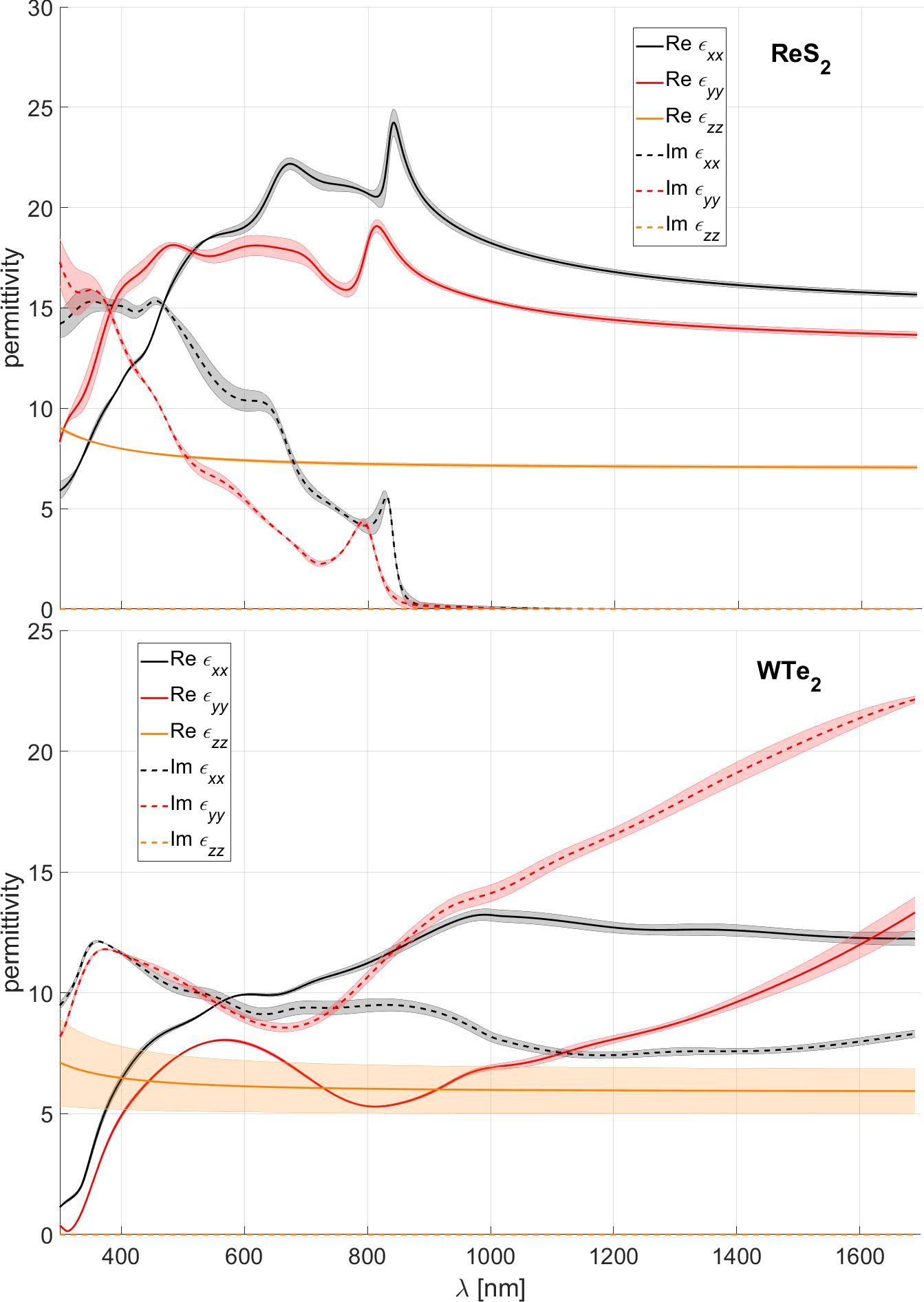}
\caption{Uncertainty of the complex permittivity for bi-anisotorpic  TMDCs.}

\label{fig:Errors-bianiso}
\end{minipage}
\hfill
\begin{minipage}[b]{.49\textwidth}
\centering
\includegraphics[width=1\textwidth]{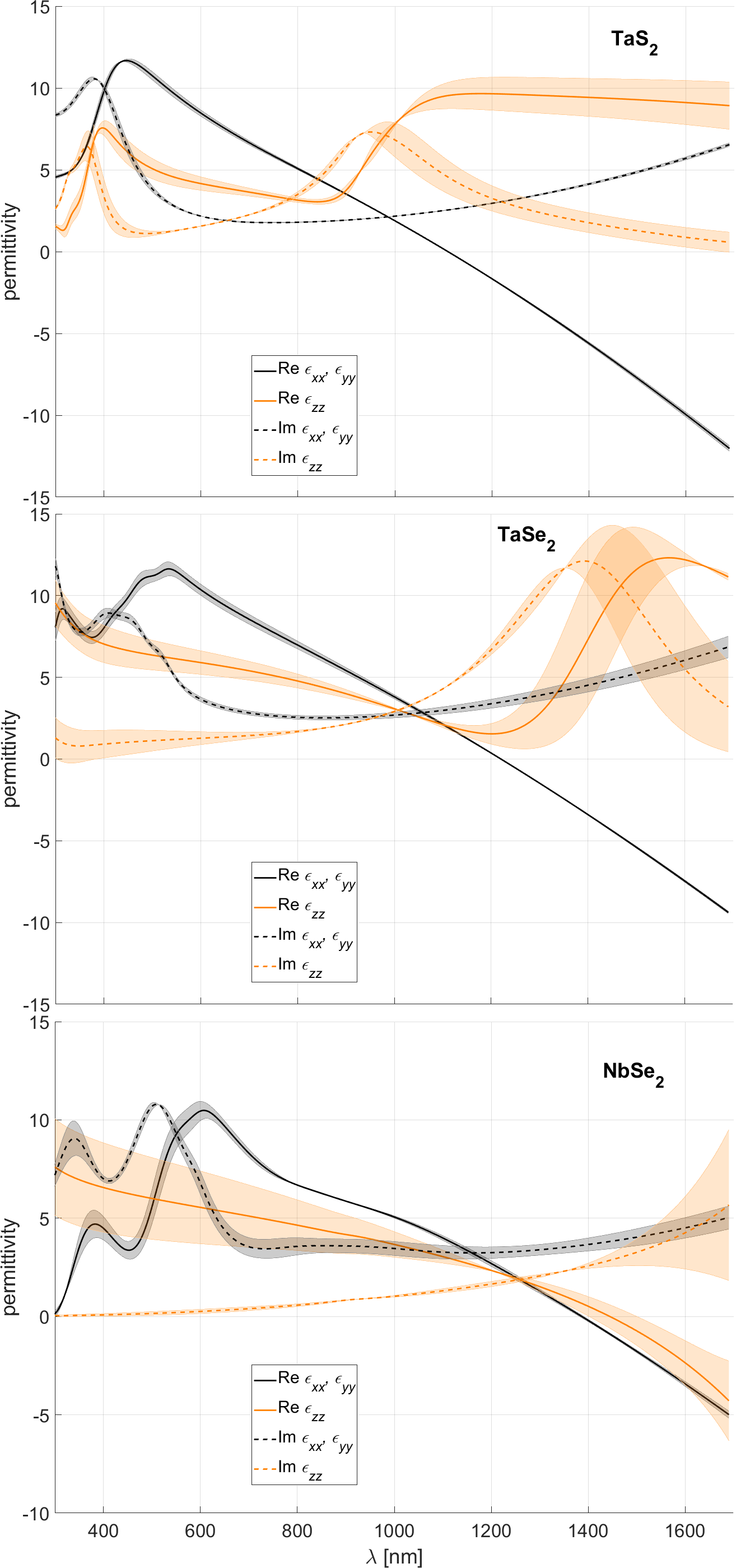}
\caption{Uncertainty of the complex permittivity for metalic TMDCs.}
\label{fig:Errors-metalic}
\end{minipage}
\end{figure}

Our analysis shows that both random and systematic errors (tested as described above) give a minor contribution to the overall uncertainty of the dielectric functions. On the other hand, the magnitude of the fit error shows the largest influence on the  models' spectral sensitivity comes from the distinct optical properties of the subsequent materials. 
For uniaxial semitransparent TMDs the permittivity errors shown in \autoref{fig:Errors-uniaxial} are negligible in the transparent regions and increase for shorter wavelengths in the absorption bands. This is a result of a lower sensitivity of ellipsometry (the method) itself due to high absorbance in the materials and an influence of the surface roughness, which plays an important role in determining the optical properties of the materials, especially in the UV region. 

In the case of bi-anisotropic materials (see \autoref{fig:Errors-bianiso}), \ce{ReS2} shows similar properties in terms of uncertainties of optical parameters to the above examples, as it is also transparent at long wavelengths. In contrast, \ce{WTe2} exhibits comparatively larger errorbars with an increase of the wavelength for the in-plane components and a pronounced uncertainty for the out-of-plane component in the entire spectral region. The latter one is attributed to the overall large absorptivity of the material in the whole spectral range. This was partially mitigated by preparation of only thin samples, however, due to the mechanical properties of \ce{WTe2} we were only able to obtain flakes with thicknesses down to $\sim$200 nm. Unfortunately, ca. \unit[200]{nm} \ce{WTe2} flakes still absorb a significant amount of light. Thus, the measured ellipsometric curves lack clear and deep  interference features despite a \ce{SiO2} layer between the \ce{WTe2} flakes and the Si substrate. These limitations of preparing \ce{WTe2} flakes led to a low sensitivity of the model to the $\varepsilon_{zz}$ component. 

Metallic TMDs show enhanced errorbars in the spectral regions with high absorptivity, as illustrated in \autoref{fig:Errors-metalic}, resulting in a decrease of the interferometic spectral features -- see \autoref{fig:mm-nbse2}--\autoref{fig:mm-tase2} -- which leads to high uncertainty of out-of-plane components. 
It is important to stress that this analysis does not evaluate the overall uncertainty of the ellipsometric technique but rather shows the sensitivity of the models resulting from unique material properties of subsequent samples